\documentclass[preprint,3p, 11pt]{elsarticle}

\usepackage{amssymb}
\usepackage{amsmath}

\usepackage{moreverb,url}
\usepackage[utf8]{inputenc}
\usepackage{longtable}
\usepackage[colorlinks,bookmarksopen,bookmarksnumbered,citecolor=blue,urlcolor=blue]{hyperref}
\usepackage{natbib}
\usepackage{multirow}
\usepackage{booktabs}
\usepackage{rotating}
\usepackage[right]{lineno}
\usepackage{setspace}
\usepackage{subfig}
\newcommand\BibTeX{{\rmfamily B\kern-.05em \textsc{i\kern-.025em b}\kern-.08em
T\kern-.1667em\lower.7ex\hbox{E}\kern-.125emX}}

\journal{}

\begin{document}

\begin{frontmatter}



\title{Quantifying Extreme Opinions on Reddit Amidst the 2023 Israeli-Palestinian Conflict}

\author[label1]{Alessio Guerra}
\author[label2]{Marcello Lepre}
\author[label3]{Oktay Karaku\c{s}\corref{1}}

\affiliation[label1]{Independent Researcher}
\affiliation[label2]{Mater Research Institute, University of Queensland, Woolloongabba QLB, 4124, Australia}
\affiliation[label3]{Cardiff University, School of Computer Science and Informatics, Abacws, Senghennydd Road, Cardiff, CF24 4AG, UK.}

\cortext[cor1]{Oktay Karaku\c{s}, School of Computer Science and Informatics, Abacws, Senghennydd Road, Cardiff, CF24 4AG, UK.}

\ead{karakuso@cardiff.ac.uk}


\begin{abstract}
This study investigates the dynamics of extreme opinions on social media during the 2023 Israeli-Palestinian conflict, utilising a comprehensive dataset of over 450,000 posts from four Reddit subreddits (\textit{r/Palestine}, \textit{r/Judaism}, \textit{r/IsraelPalestine}, and \textit{r/worldnews}). A lexicon-based, unsupervised methodology was developed to measure "extreme opinions" by considering factors such as anger, polarity, and subjectivity. The analysis identifies significant peaks in extremism scores that correspond to pivotal real-life events, such as the IDF's bombings of Al Quds Hospital and the Jabalia Refugee Camp, and the end of a ceasefire following a terrorist attack. Additionally, this study explores the distribution and correlation of these scores across different subreddits and over time, providing insights into the propagation of polarised sentiments in response to conflict events. By examining the quantitative effects of each score on extremism and analysing word cloud similarities through Jaccard indices, the research offers a nuanced understanding of the factors driving extreme online opinions. This approach underscores the potential of social media analytics in capturing the complex interplay between real-world events and online discourse, while also highlighting the limitations and challenges of measuring extremism in social media contexts.
\end{abstract}




\end{frontmatter}



\section{Introduction}
The Israeli-Palestinian conflict, one of the most long-lasting and contentious geopolitical issues, has remained active for over a century, marked by episodes of violence, negotiation, and temporary ceasefires. This conflict has far-reaching implications not only for the region but also for international politics, influencing global diplomatic relations and motivating strong emotions worldwide. In recent years, the rise of social media platforms has transformed the landscape of public communications, providing a space where individuals express their views and engage in debates on contentious issues, including the Israeli-Palestinian \citep{liyih2024sentiment,chen2024isamasred} and/or Russo-Ukrainian \citep{tschirky2024azovsteel,aviv2023russian} conflicts.

Social media platforms like Reddit offer a unique perspective through which to examine public opinion, especially regarding polarising topics. Reddit, with its diverse user base and structure of subreddits dedicated to specific interests and discussions, serves as a valuable resource for analysing public sentiment. The anonymity and relative freedom provided by such platforms often result in the expression of extreme opinions, which might be less visible in traditional media or face-to-face interactions. These extreme opinions can significantly influence public perception and the overall narrative surrounding the conflict.

This study aims to investigate the nature and dynamics of extreme opinions related to the Israeli-Palestinian conflict as expressed on Reddit. By analysing posts and comments over time, we seek to understand how these opinions evolve, what factors contribute to their extremity, and how they reflect broader societal trends. Our focus on extreme opinions is crucial, as these can provide insights into the underlying tensions and grievances that moderate opinions might not reveal. Moreover, understanding the shifts in extreme opinions can help in identifying potential flash-points for conflict escalation or opportunities for peace-building.

In this paper, we employ a combination of natural language processing (NLP) techniques, sentiment analysis and text processing to systematically study the opinion flow on Reddit. We will analyse the frequency, intensity, and nature of extreme opinions, exploring the context in which these opinions are expressed and how they change over time. Additionally, we aim to identify key events or triggers that may correlate with significant shifts in public sentiment.

Throughout the paper, we define ``\emph{extreme opinions}" as expressions that exhibit high emotional intensity, polarizing language, or an unusually strong stance relative to the discourse within the observed subreddits. This definition is operationalized through computational measures that capture both sentiment polarity and emotional valence. While ``\emph{extreme expression}" may not always equate to ``\emph{extremist opinions}", we argue that high emotional intensity often correlates with polarizing or inflammatory language, which can exacerbate divisive discourse. To strengthen this distinction, we have included specific examples of posts from the dataset. For instance, a highly negatively valenced expression like ``This act is pure evil and unforgivable" may differ in context and intent from a moderate expression of disapproval such as ``This action is disappointing and should be condemned." The former aligns more closely with our definition of ``extreme opinions." By including examples and contextualizing the concept within the broader literature on extremism, we aim to provide clarity on the limitations and scope of our operational definition.

The insights gained from this study hold significant implications for digital media and computational journalism, providing a deeper understanding of how social media platforms amplify and shape public discourse on complex geopolitical conflicts like the Israeli-Palestinian issue. By systematically analysing extreme opinions on Reddit, we aim to uncover patterns and dynamics that traditional media often overlooks. This research not only contributes to the academic understanding of online discourse but also informs computational journalism practices, offering methodologies to track and analyse sentiment and opinion dynamics in real-time. Such insights are crucial for developing more informed digital communication strategies, enhancing media literacy, and promoting constructive dialogue amidst polarising narratives.

\textbf{Research Questions:}  
This paper seeks to answer the following research questions:

\begin{enumerate}
  \item \textbf{RQ1:} How do patterns of extreme sentiment evolve on Reddit during key events in the 2023 Israeli-Palestinian conflict?
  \item \textbf{RQ2:} What are the differences in sentiment extremity across thematically distinct subreddits (e.g., cultural, geopolitical, general news) during the conflict?
  \item \textbf{RQ3:} How effective are lexicon-based computational models in identifying extreme opinions in polarized Reddit discourse?
\end{enumerate}

In the sequel, we first present the literature pieces and academic background about the topic in Section~\nameref{sec:related}. Methodological details of the presented work are given in Section~\nameref{sec:meth} whilst Section~\nameref{sec:exp} presents the experimental analysis and results. Section~\nameref{sec:critical} concludes the paper with a critical discussion of the strengths and limitations of the proposed technique, as well as future work directions and concluding remarks.

\section{Related Works}\label{sec:related}
\subsection{Media, Technology and Conflicts}
For the past several decades, mainstream media (especially television, radio, and newspapers) have played a significant role in shaping conflicts, guiding transitions from war to peace, contributing to violence, and affecting post-conflict reconstruction efforts \citep{schoemaker2014media}. It has become evident that the media can influence foreign policy by eliciting emotional responses from their audiences, which in turn puts pressure on governments to take action. For instance, \citet{hawkins2002other} has argued that media agendas impact a wide array of policy initiatives and that insufficient media coverage can lead to a lack of policy development. 

In the digital age, social media's influence on conflict scenarios has become highly significant, potentially surpassing that of traditional media. Platforms such as Reddit, Twitter, and Facebook have democratised information dissemination, enabling individuals to share real-time updates and unfiltered viewpoints on ongoing conflicts. Social media not only extends the reach of traditional media but also introduces new dynamics into public discourse, often intensifying polarisation and spreading misinformation. Consequently, understanding social media's role in conflicts—particularly how it shapes public opinion and influences policy responses—is crucial.


The aforementioned intersection of social media analysis, sentiment analysis, and conflict studies has attracted substantial academic attention in recent years. Researchers have utilised these platforms to explore the dynamics of opinion formation, the spread of misinformation, and the polarisation of views. \citet{zeitzoff2017social} demonstrated compelling evidence that social media has been deliberately utilised to influence and manipulate audiences. For example, ISIS leveraged social media to target enemies and recruit followers, Russia engaged in cyberwarfare and spread misinformation, and Trump used Twitter as a strategic tool during his 2016 presidential campaign. \citet{vera2018sentiment} analyse sentiments and opinions on Twitter regarding the Colombian post-conflict period that began in 2016. The study collected 250 tweets from Colombians and 250 tweets from foreigners to compare perceptions. The analysis revealed that foreigners' tweets were significantly more positive, with 60\% expressing positive sentiments, compared to only 20\% of Colombians' tweets. \citet{ozturk2018sentiment} conduct the first comparative study to examine sentiments regarding the Syrian refugee crisis. They carried out an extensive sentiment analysis on related Turkish and English tweets. The Turkish data revealed a more balanced distribution with generally positive sentiments toward refugees, whereas the English-speaking community expressed significantly more neutral and negative opinions.

Having discussed several examples of conflicts and their corresponding social media sentiment analyses, it is evident that sentiment analysis serves as a powerful tool for understanding public opinion in various geopolitical contexts. Continuing this examination, we will explore additional case studies that further illustrate the diverse applications and insights provided by sentiment analysis in conflict scenarios. The study by \citet{aslan2023deep} applies deep learning techniques to analyse public sentiment on the Ukraine-Russia war using a large dataset of geo-tagged tweets. It employs topic modelling and sentiment analysis, including the Valence Aware Dictionary for Sentiment Reasoner (VADER) tool, and introduces a new sentiment classification model, MF-CNN-BiLSTM. Experimental results confirm its successful performance in sentiment analysis tasks. \citet{akpatsa2022sentiment} analysed 300K+ tweets about the US troop withdrawal from Afghanistan (August 11-27, 2021) using text mining techniques. Findings showed predominantly negative social media reactions, with diverse topics discussed related to the crisis fallout. Machine learning classifiers, particularly support vector machines (SVM), achieved high accuracy (0.83) and demonstrated significant precision, recall, and f1-score metrics. \citet{guerra2023sentiment} present a novel lexicon-based unsupervised sentiment analysis method to gauge "hope" and "fear" during the 2022 Ukrainian-Russian Conflict using Reddit.com as the data source. Daily scraping of the top 50 posts from six relevant subreddits generated a unique dataset spanning nearly the conflict's first three months. Analyses included assessing public interest, calculating Hope/Fear scores, and exploring interactions with stock prices. Results showed a marked decline in hope following significant losses like Azovstal in Mariupol and Severodonetsk, with fluctuations also observed after non-military events such as Eurovision and football games.

In parallel with the efforts mentioned above in shaping the new media for public discourse studies, social media platforms have also become productive grounds for the increase of extreme opinions and the polarisation of viewpoints. \citet{del2016spreading}, by analysing Facebook posts as scientific and conspiracy groups, demonstrate that social media allows for the rapid dissemination of unsubstantiated rumours and conspiracy theories that often elicit rapid, large, but naive social responses such as the unfortunate case of Jade Helm 15 where a simple military exercise turned out to be perceived as the beginning of a new civil war in the United States. \citet{matakos2017measuring} study measuring and reducing the polarising opinions in social media using a standard opinion formation model. They define a polarisation index, which quantifies the polarisation observed in the network. 

\subsection{Advances in Social Media Text Analytics}
As the prevalence of extreme opinions and polarisation on social media becomes increasingly evident, leveraging big data through advanced text analysis techniques such as sentiment analysis and topic modelling offers invaluable insights into the underlying dynamics of these phenomena. Sentiment analysis involves employing techniques such as natural language processing, text analysis, and mining to extract and interpret subjective information from various sources, including written text and speech \citep{MEDHAT20141093, liu2020sentiment}. This field encompasses tasks like opinion mining, sentiment mining, emotion analysis, and more \citep{liu2020sentiment, nasukawa2003sentiment, dave2003mining}. Text data mining, which involves extracting information from text-based data sources, serves diverse purposes and utilises techniques such as topic modelling and sentiment analysis \citep{hearst1999untangling, rehurek2010software, feldman2013techniques}. Text-related sentiment analysis is versatile, supporting applications ranging from assessing psychological disorders \citep{zucco2017sentiment} and analysing human behaviour during events like the World Cup \citep{yu2015world}, to detecting general emotions \citep{peng2021survey}, exploring gender differences \citep{thelwall2010data}, predicting stock market trends \citep{pagolu2016sentiment}, and assessing investor sentiment via social media posts \citep{10.3389/frai.2022.884699}.

Social media has naturally emerged as a primary source for text mining and sentiment analysis due to its expansive user base and broad reach across communities. Twitter stands out as a prominent example where sentiment analysis has been extensively applied \citep{yu2015world, 10.3389/frai.2022.884699, hu2013unsupervised, qi2023sentiment}. Moreover, other popular platforms like Facebook \citep{ortigosa2014sentiment,laabar2024multi}, Reddit \citep{guerra2023sentiment, melton2021public}, MySpace \citep{thelwall2010data}, Telegram \citep{ptaszek2024war}, and even YouTube comments \citep{muinao2024youtube,yuliansyah2024sentiment} have also been utilised for sentiment analysis purposes.

\subsection{On Israeli-Palestinian Conflict}
Previously, we touched base on the importance of sentiment analysis of social media for understanding conflicts and discourses. Building on these advancements in computational journalism, sentiment analysis provides a valuable tool for examining the complex and often polarised discourse surrounding the Israeli-Palestinian conflict on social media platforms. In \citep{schrodt2011forecasting}, the Latent Dirichlet Allocation (LDA)-based method can forecast interaction patterns and conflict measures by using event data for 29 Asian countries (1998-2010) and the Israel-Palestine and Israel-Lebanon dyads (1979-2009). LDA combined with a logistic model predicts outcomes with 60-70\% accuracy and improves classification sensitivity over simple logistic models. However, a supervised LDA version does not significantly improve results and shows some issues. \citet{chambers-etal-2015-identifying} use sentiment analysis to detect international relations from social media, analysing hundreds of country pairs over 17 months. It finds a strong correlation (\(\rho = 0.8\)) between sentiment scores and public opinion polls, validating the approach beyond simple positive or negative classifications. The methodology could automate expensive human polls. Positive sentiment analysis showed high accuracy, while negative sentiment accuracy was 68\%, improving to 81\% with higher observed counts, noting the under-representation of smaller nation conflicts on Twitter. \citet{aouragh2016palestine} provides foundational insights into the role of digital media in shaping narratives around the Israeli-Palestinian conflict. In viewing the Internet as a political tool, enabler of community, and means of news dissemination, Aouragh's study holds lessons for analysing its role throughout the rest of the region \citep{york2012palestine}.

Furthermore, \citet{al2019multi} analyse Twitter data to measure public opinion on the Palestinian-Israeli conflict using a novel two-level data analysis model. The country-level analysis examines overall attitudes by identifying countries generating the most tweets, measuring each country's friendliness towards Palestine, and analysing sentiment changes over time. The individual-level analysis focuses on the activity and background of individuals, particularly opinion leaders and ethnic groups, concerning country attitudes. The study employs a logistic regression-based ``political sentiment" classifier, which outperforms standard classifiers and provides detailed insights. \citet{matalon2021using} investigated "opinion diffusion," where users often reverse a message's content before sharing it, thereby creating a contrasting view. By analysing 716,000 Tweets on the Israeli-Palestinian conflict, the study identified 7147 Source-Quote pairs. Utilising a Random Forest model with NLP features, they predicted Opinion Inversion (O.I.) with an ROC-AUC of 0.83. The findings indicate that 80\% of the factors contributing to O.I. are related to the sentiment of the original message, underscoring O.I.'s significant role in political communication on social media.

The current research by \citet{arapostathis2023archiving} presents a method for archiving social media discussions over time and space using a dataset of 1.4 million Greek tweets from January 2015 to April 2016. LDA models were generated to extract 1,780 topics, classified into three categories: Refugee, War, and Irrelevant. Named Entity Recognition (NER) was used to extract and geocode approximately 940,700 geolocations. The study produced figures and maps showing the frequency of geolocations in Syria and Iraq. The research highlights the effectiveness of using transformers for geoparsing and provides insights into LDA model parametrisation, revealing dominant discussions and their geographic focus. In another recent study, to address the challenge of analysing diverse public opinions on YouTube regarding the Hamas-Israel war, a deep learning-based sentiment analysis model was developed by \citet{liyih2024sentiment}. A total of 24,360 comments were collected from popular YouTube news channels such as BBC, WION, and Aljazeera. Comments are labelled as positive, negative, or neutral. Textual comments were preprocessed and features were extracted using Word2vec, FastText, and GloVe. The SMOTE data balancing technique was applied. Classification algorithms LSTM, Bi-LSTM, GRU, and a Hybrid of CNN and Bi-LSTM were tested, with the Hybrid model and Word2vec achieving the highest classification accuracy of 95.73\%.

\subsection{Research Gaps}

Despite substantial research in the field of sentiment analysis, there are notable gaps in understanding the dynamics of extreme opinions within the context of geopolitical conflicts, especially in social media discourse \citep{wadhwani2023sentiment, renhoran2024public}. While many studies examine sentiment on social media platforms, they tend to focus on general sentiment analysis and often neglect the deeper, more polarizing opinions that are essential for understanding the intensity of conflict-related discourse \citep{shukla2021sentiment}. Furthermore, there is limited research on how extreme opinions evolve over time in response to key events, which is particularly relevant in ongoing and fast-developing geopolitical conflicts \citep{rad2018exposure, sentiment2023framework}.

Existing studies on online sentiment typically analyse static or aggregated views of public sentiment, rather than exploring the shifts and triggers that drive extreme opinions within specific subgroups of online communities \citep{tornberg2021modeling, sasikumar2023sentiment}. Additionally, the identification of extreme opinions, particularly those expressed on platforms like Reddit, where anonymity can lead to more emotionally charged language, remains an underexplored area \citep{sasikumar2023sentiment}. There is also a lack of detailed comparison across different types of subreddits (e.g., political, cultural, or general news subreddits), which limits the understanding of how extreme opinions manifest in various thematic contexts \citep{sentiment2023framework, de2021no}.

This paper addresses these gaps by focusing on the dynamics of extreme opinions within the context of the Israeli-Palestinian conflict on Reddit. It investigates how extreme sentiment evolves over time, particularly in response to key events in the conflict. The study also examines sentiment extremity across different subreddits, helping to highlight contextual differences in the expression of extreme opinions. Furthermore, the paper explores the effectiveness of lexicon-based computational models in identifying these extreme opinions, an area that has received little attention in existing literature.

\section{Methodology}\label{sec:meth}
This paper delves into the 2023 Israeli-Palestinian conflict by analysing polarising extreme opinions using a comprehensive dataset of social media posts collected from Reddit. The study focuses on four subreddits: \textit{r/Palestine}, \textit{r/Judaism}, \textit{r/IsraelPalestine}, and \textit{r/worldnews}, amassing a total of over 450,000 comments. Our proposed approach employs a lexicon-based unsupervised method to quantify "extreme opinions." This methodology involves several stages, including data preprocessing, sentiment analysis, and the calculation of extremism scores. 

In the sequel, the methodology is explained step-by-step, beginning with details of the Reddit dataset, followed by pre-processing stages, extreme opinion measurement, and statistical analysis.

\subsection{Reddit Data Details \& Preparation}
Reddit was selected for this study due to its unique structure, which facilitates detailed discussions within specific communities. Unlike other social media platforms such as Twitter, where individual accounts drive content visibility, Reddit is organised around \textit{subreddits}, which are niche communities dedicated to various topics. This structure allows for a more focused and in-depth analysis of topics, making it easier to examine how extreme opinions are formed and evolve within specific discourse spaces. Furthermore, Reddit's emphasis on anonymity and the relatively unfiltered nature of many subreddits foster a space where users can express polarized or extreme opinions that may not be as prevalent in mainstream media. This makes Reddit an ideal platform to investigate public sentiment on contentious issues such as the Israeli-Palestinian conflict \citep{guerra2023sentiment}.

For this study, data was collected using the Reddit API, which provides access to posts and comments from publicly available subreddits. To gather the necessary data, a Python script was developed to scrape the top 50 posts and their comments from each selected subreddit on a daily basis. The data collection process began on October 7, 2023, and concluded on January 3, 2024, with data being collected daily around mid-day UK time. 

The following four subreddits were chosen for this analysis:

\begin{itemize}
    \item \textit{r/IsraelPalestine} - This subreddit is dedicated to discussions about the Israeli-Palestinian conflict. It serves as a primary source of information and opinion, attracting both general and deeply engaged discussions from users with varying perspectives on the conflict. Given its focus, it provides a rich dataset of opinions directly related to the topic under investigation.
    \item \textit{r/worldnews} - As a large and general subreddit, \textit{r/worldnews} covers a wide range of global events, including the Israeli-Palestinian conflict. While this subreddit features broader international perspectives, it is valuable for capturing public reactions to major news events related to the conflict and understanding how they influence the wider discourse.
    \item \textit{r/Palestine} - This subreddit focuses specifically on the Palestinian perspective, providing insights into the views and narratives shared by users who may identify with or have a strong interest in Palestinian issues. The focus on this community helps in understanding the views from one side of the conflict, which is crucial for capturing the range of extreme opinions.
    \item \textit{r/Judaism} - This subreddit provides insight into Jewish perspectives on various issues, including the Israeli-Palestinian conflict. Given the central role of religion and identity in the conflict, this subreddit offers valuable context for understanding how religious and cultural identities shape extreme opinions in online discussions.
\end{itemize}

These subreddits were selected based on their relevance to the Israeli-Palestinian conflict and their potential to provide diverse perspectives. By including subreddits representing multiple sides of the issue, we ensure that the dataset reflects a wide spectrum of opinions, including those that may be more extreme or polarized. This diverse selection also allows us to track how different user communities engage with the same events and topics.

Data collection was performed using a Python script that retrieved the top 50 posts and their associated comments from each subreddit on a daily basis. The script combined newly gathered submissions with previously collected data and removed duplicates based on submission IDs. The total number of comments collected for each subreddit is presented in Table \ref{tab:subr}, and additional information for each submission is shown in Table \ref{tab:info}.

\begin{table}[htbp]
    \centering
    \caption{Utilised subreddits and number of data samples for each.}\label{tab:subr}
    \begin{tabular}{p{6cm}p{5.7cm}}
    \toprule
     \textbf{Subreddit}    &  \# \textbf{of comments}\\
     \toprule
     \textit{r/IsraelPalestine}  &  253,888\\ \hline
    \textit{r/Judaism}              &3,839\\\hline
    \textit{r/Palestine}           &61,410\\\hline
    \textit{r/worldnews}          &160,793\\\bottomrule
    \textbf{Total}                 & 479,930\\\bottomrule
    \end{tabular}  
\end{table}

\begin{table}[htbp]
    \centering
    \caption{Some details for each data collected sample.}\label{tab:info}
    \begin{tabular}{p{3.7cm}p{12cm}}
    \toprule
     \textbf{Column}    &  \textbf{Description}\\
     \toprule
    title  & (only for posts) the title of the post\\\hline
    text & the actual content of the submission\\\hline
    upvotes & a method by which users can show their approval/support for a post\\\hline
    flair & categorisation of the post by the author\\\hline
    subreddit & subreddit name that each comment belongs to\\\bottomrule
    \end{tabular}  
\end{table}

Data preparation for the extremism measurement followed a similar approach to \citep{guerra2023sentiment}, including the cleaning of irrelevant comments, lowercasing, and counting upvotes and word counts. Irrelevant posts, such as those in \textit{r/worldnews} that were not directly related to the conflict, were filtered out by retaining only posts with relevant flairs like “Israel/Palestine” and “Israel Megathread.” Since flairs are assigned only to posts, the structure of Reddit allows for tracing comments back to their original posts using submission IDs, enabling the identification and removal of irrelevant submissions.

\subsection{Measuring Extreme Opinions}
As previously discussed, this paper's main purpose - measuring and quantifying extreme opinions on social media posts - is a critical endeavour for understanding the dynamics of online discourse, especially in contentious contexts such as the Israeli-Palestinian conflict. Extreme opinions, often characterised by intense, polarised, and sometimes inflammatory language, can significantly influence public perception and exacerbate community divisions. Researchers can identify and quantify the presence and intensity of extreme viewpoints by analysing the language, tone, and sentiment expressed in social media posts. In this section, we detail our proposed technique for measuring extreme opinions, leveraging advanced NLP methods to systematically assess and quantify the extremity of opinions expressed in Reddit comments related to the Israeli-Palestinian conflict.

Preprocessed Reddit comments are utilised through several stages to create the extremism measure:

\textbf{Anger Score ($a$):} Using the \textit{nltk} module’s VADER Sentiment tools \citep{hutto2014vader}, negatively polarised text was identified as anger-reflecting comments, resulting in the anger score. This score ranges from 0.0 to 1.0, where higher values indicate higher levels of anger.

\textbf{Polarity Score ($p$):} Employing a different sentiment analysis approach, the text of a post or comment receives a polarity score ranging from -1 to 1, with -1 indicating very negative sentiment and 1 indicating very positive sentiment. This score is extracted using the \textit{sentiment.polarity} method from the \textit{TextBlob} Python module.

\textbf{Subjectivity Score ($s$):} The \textit{sentiment.subjectivity} method from the same \textit{TextBlob} module helps determine if the author is stating facts or expressing opinions. Subjectivity scores range from 0 (objective) to 1 (subjective).

\textbf{On Justification for Combining VADER and \textit{TextBlob}:} As stated above, our methodology integrates VADER and \textit{TextBlob} to derive a composite measure of sentiment, capturing both polarity and emotional intensity. VADER is particularly effective at handling short social media texts with emotive language \cite{hutto2014vader}, whilst \textit{TextBlob} excels in determining subjectivity \cite{Hazarika2020, Tinnalur2021}. By combining these, we aim to balance nuanced interpretations of sentiment with computational robustness. This approach is supported by prior studies that highlight the complementary nature of these tools in analysing social media discourse \cite{Gupta2024,Barik2024}.

\textbf{Weighting for Length ($(\cdot)^l$):} Dictionary-based sentiment analysis often favours longer texts due to the higher word count, increasing the chances of finding relevant words and potentially inflating scores. To address this, we weigh the score of each text by its number of words.

\textbf{Weighting for Upvotes ($(\cdot)^u$):} Some opinions are more popular than others, and on Reddit, popularity can be gauged by the number of upvotes. To better reflect public opinion, we weighted the score by the number of upvotes. A comment viewed by 100 people and upvoted by 10 (10\%) has a higher score than a comment viewed by 10 people and upvoted by 5 (50\%). To account for this, the number of upvotes was weighted against the number of comments on a post to determine its relative popularity.


\textbf{Extremism Score ($\chi$):} Combining all the stages above is crucial for obtaining an informative measure of extreme opinions. Our proposed extremism score is calculated as follows:

\begin{equation}
\chi = a \cdot |p|
\end{equation}

This extremism expression has several advantages. Using both anger and polarity scores captures the intensity of the emotion and the sentiment direction (positive or negative). This approach helps identify posts that are not only polarised but also emotionally charged. Taking the absolute value of the polarity ensures that both strongly positive and strongly negative opinions are considered extreme, which is crucial for identifying polarisation.

It is clear from the expression that it does not initially account for the subjectivity score of posts. As previously mentioned, the subjectivity score, which measures the extent to which a statement is opinion-based rather than fact-based, can add another dimension to our measure of extreme opinions. Including subjectivity helps differentiate between emotionally charged factual statements and highly subjective opinions. Therefore, our definition of the extremism score takes the form of:

\begin{equation}
\chi = a \cdot |p| \cdot s
\end{equation}

Here, the subjectivity score amplifies the extreme opinion score, making subjective posts stand out more. Finally, this expression ensures that posts with high anger, high polarity, and high subjectivity will have the highest extreme opinion scores. On the other hand, considering all three multipliers above are values between 0 and 1, it is possible to have relatively small extremism scores. This might make the interpretability of the score challenging. At this point, we propose normalising the extremism score by the product of the mean values of each component (anger, absolute polarity, and subjectivity). We believe that this enhances the robustness of the measure by reducing the influence of outliers, thus providing a more reliable and consistent assessment of extreme opinions. This approach scales the raw extremism score relative to the average values in the dataset, allowing for more meaningful comparisons across different datasets or time periods. By applying this, we achieve a balanced and contextually relevant measure of extremism. Hence, our extremism score is defined as:

\begin{align}
    \chi_{\text{norm}} = \frac{a \cdot |p| \cdot s}{\overline{a} \cdot \overline{|p|} \cdot \overline{s}}
\end{align}

\begin{figure}[htbp!]
\centering
\includegraphics[width=0.9\linewidth]{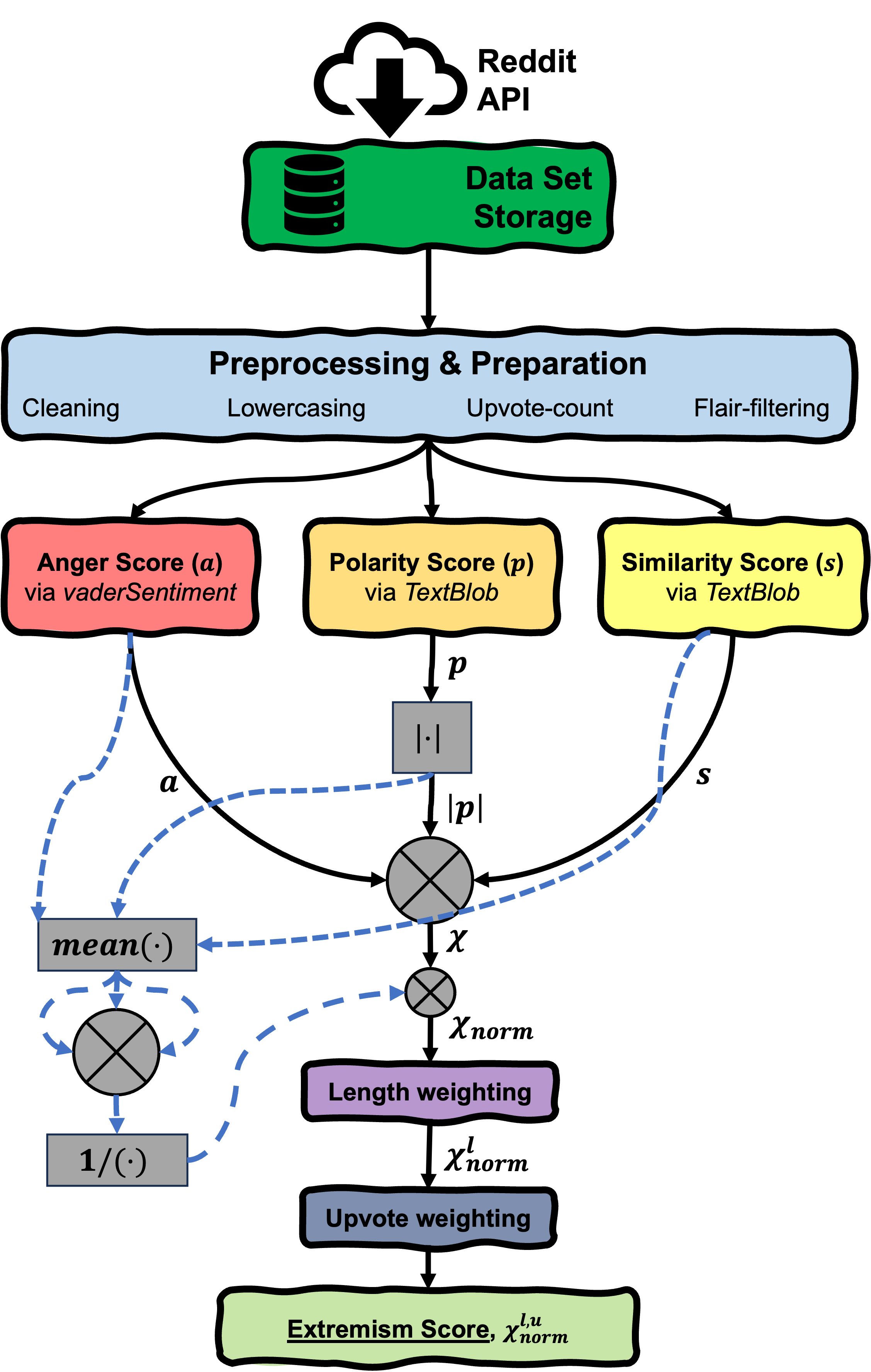}
\caption{A flow diagram to demonstrate the proposed extremism score calculation stages.}
\label{fig:flow}
\end{figure}

Finally, the normalised extremism score is weighted by length and upvotes as suggested above. The weighting stages of the normalised extremism score are given as
\begin{align}
    \chi^l_{\text{norm}} &= \dfrac{\chi_{\text{norm}}}{\log\left(L+1\right)}.\\
    \chi^{l,u}_{\text{norm}} &= \chi^l_{\text{norm}} \times \left(1 + \dfrac{u}{n} \right).
\end{align}
where $u$ is the number of upvotes, $L$ is the number of words and $n$ refers to the number of submissions on that specific post. The general procedure for calculating the extremism score and the details of each variable are illustrated in Figure \ref{fig:flow}.


\section{Experimental Analysis}\label{sec:exp}
To comprehensively evaluate the proposed extremism score and its implications, we conducted a series of experimental analyses, each designed to examine different facets of our data and methodology:

\begin{enumerate}
    \item \textbf{Statistical Analysis of Scores:} Following the validation of the proposed extremism score, by utilising the collected Reddit data, we performed a detailed statistical analysis of the derived scores (anger, polarity, and extremism score). 
    
    \item \textbf{Score Validation on a Manually Created Dataset:} First, the developed extremism score and its components were validated on a 50-sample manually labelled data set.
    
    \item \textbf{Interest Over Time:} We then analysed the interest in the topic over time by tracking the frequency and volume of relevant social media posts. 

    \item \textbf{Analysing Extreme Opinions:} In this section, we delved into the nature of extreme opinions themselves. 

    \item \textbf{IDF vs. Hamas in Sentiments:} We specifically compared sentiments directed towards the IDF (Israel Defense Forces) and Hamas to highlight differences in public opinion. 

    \item \textbf{Quantitative Effects of Each Score on Extremism Score:} Next, we quantitatively assessed the impact of each component score (anger, polarity, and subjectivity) on the extremism score. 

    \item \textbf{Corpus Similarity:} We analysed corpus similarity to examine the linguistic patterns and common themes across the dataset. 

    \item \textbf{Chronological Validation:} Finally, we conducted chronological validation to ensure the consistency and reliability of our extremism score over time. 
\end{enumerate}

\subsection{Statistical Analysis of Scores}
In this subsection, we conduct a statistical analysis of the anger, absolute polarity, inverted subjectivity, and both the normalised and weighted normalised extremism scores, focusing on Pearson correlation and descriptive statistics. This analysis allows us to understand the distribution, central tendency, and variance of each score, offering insights into their characteristics and interrelationships. Table \ref{tab:stats} provides details such as correlation coefficients, means, and medians for each score.
 
\begin{table}[htbp]
  \centering
  \caption{Statistical Analysis of Scores}
    \begin{tabular}{clp{2.25cm}p{2.25cm}p{2.25cm}p{2.25cm}p{2.25cm}}
    \toprule
          &       & $a$ & $|p|$ & $s$ & $\chi_{\text{norm}}$ & $\chi_{\text{norm}}^{l,u}$ \\ \toprule
    \multirow{5}[0]{*}{\begin{turn}{60}Pearson correlation\end{turn}} & $a$ & 1.000 & 0.140 & 0.188 & 0.496 & 0.365 \\
          & $|p|$ & 0.140 & 1.000 & 0.603 & 0.555 & 0.373 \\
          & $s$ & 0.188 & 0.603 & 1.000 & 0.380 & 0.246 \\
          & $\chi_{\text{norm}}$ & 0.496 & 0.555 & 0.380 & 1.000 & 0.766 \\
          & $\chi_{\text{norm}}^{l,u}$ & 0.365 & 0.373 & 0.246 & 0.766 & 1.000 \\
          &       &       &       &       &       &  \\ \bottomrule \bottomrule
    \multirow{7}[0]{*}{\begin{turn}{60}Descriptive Statistics\end{turn}} & mean  & 0.120 &	0.155	&0.391	&1.934	&0.405 \\
          & std   & 0.130	& 0.186	& 0.280	& 6.189	& 2.145 \\
          & min   & 0.000 & 0.000 & 0.000 & 0.000 & 0.000 \\
          & 25\%  & 0.000 & 0.000 & 0.125 & 0.000 & 0.000 \\
          & 50\%  & 0.097	& 0.100	& 0.420	& 0.172	& 0.025 \\
          & 75\%  & 0.188	& 0.225	& 0.574	& 1.332	& 0.214 \\
          & max   & 1.000 & 1.000 & 1.000 & 137.624 &	516.889 \\
          \bottomrule
    \end{tabular}%
  \label{tab:stats}%
\end{table}%

The correlation analysis in the Table \ref{tab:stats} presents the Pearson correlation coefficients between the various scores where the key observations include:

\textbf{Anger - $a$} shows a moderate positive correlation with both $\chi_{\text{norm}}$ (0.496) and $\chi_{\text{norm}}^{l,u}$ (0.365). This indicates that higher anger scores are associated with higher extremism scores, supporting the hypothesis that anger is a significant component of extreme opinions. Furthermore, \textbf{Absolute Polarity - $|p|$} has a higher positive correlation with $\chi_{\text{norm}}$ (0.555) and $\chi_{\text{norm}}^{l,u}$ (0.373). This suggests that the intensity of sentiment (whether positive or negative) contributes to extremism, even stronger than anger. Moreover, \textbf{Subjectivity - $s$} shows a positive correlation with all other scores, with particularly strong relationship with $|p|$ (0.603). This indicates that more subjective comments tend to have higher polarity scores. Lastly, \textbf{Raw Normalised Extremism - $\chi_{\text{norm}}$} is highly correlated with the weighted final extremism score - $\chi_{\text{norm}}^{l,u}$ (0.766), demonstrating that the weighting process maintains the relative relationships while adjusting for factors such as length and upvotes. Consequently, we can conclude that both anger and polarity are significant contributors to the extremism score, while subjectivity plays a more complex role, highly related to polarity and less directly linked to extremism.

Examining the descriptive statistics results in Table \ref{tab:stats}, the mean values indicate that, on average, the scores are relatively low whilst the standard deviations show that $\chi_{\text{norm}}$ and $\chi_{\text{norm}}^{l,u}$ have higher variability (6.189 and 2.145, respectively) compared to the other scores, suggesting greater dispersion in extremism measures. Minimum values are zero for all scores, indicating the presence of comments with no anger, polarity, subjectivity, or extremism. The 25th percentile values are also zero except for $s$, implying that at least 25\% of the comments have very low or no scores in these dimensions. For $s$, the 25th percentile is 0.125, indicating a higher baseline level of subjectivity. The median values (50th percentile) show that half of the comments have scores below these values which are relatively small. The 75th percentile values highlight that the upper quartile of the data has higher scores, particularly for $s$ at 0.574, and $\chi_{\text{norm}}$ at 1.332.

\subsection{Validation of Scores}
To evaluate the proposed methodology effectively, a labelled validation dataset was generated specifically for the task. This dataset was synthetically designed to simulate Reddit-like posts, representing a mix of subreddits with diverse thematic contexts (e.g., political, cultural, and general news). Each post was manually annotated to indicate its sentiment intensity, polarity, subjectivity, and extremity. The extremism scores were calculated purely based on linguistic and contextual features derived from the text, excluding any engagement metrics like upvotes or number of submissions. Six example lines of the validation set, along with their labels and extremism scores, are presented in Table \ref{tab:valdata}.

\begin{table}[htbp]
  \centering
  \caption{Six example posts from the manually labelled validation data set with their labels and extremism scores.}
    \begin{tabular}{p{0.75cm}p{3cm}p{7.5cm}p{2cm}p{1.25cm}}
    \toprule
    Post ID & Subreddit & Text  & Label & $\chi^l_{\text{norm}}$ \\
    \toprule
    1     & r/Palestine & This is genocide, and the world is blind to it! Shame on the international community! & Extreme & 0.636 \\\hline
    2     & r/worldnews & Another tragic escalation in this conflict. Innocent lives are lost again. & Moderate & 0.234 \\\hline
    3     & r/Judaism & Peace must prevail. Violence is never the answer, no matter who starts it. & Neutral & 0.000 \\\hline
    12    & r/IsraelPalestine & How can anyone justify these actions? It's pure evil, and the world must take a stand against this barbarity. & Extreme & 0.673 \\\hline
    18    & r/Judaism & Israel has a right to defend itself, but the loss of civilian lives is unacceptable. & Moderate & 0.613 \\\hline
    37    & r/worldnews & The conflict has sparked widespread protests in major cities worldwide. & Neutral & 0.087 \\\bottomrule
    \end{tabular}%
  \label{tab:valdata}%
\end{table}%

The evaluation of the labelled validation dataset, as exemplified in Table \ref{tab:valdata}, highlights several key points. Firstly, the extremism scores align well with the qualitative labels assigned to the posts. For instance, posts labelled as "Extreme" (e.g., IDs 1 and 12) exhibit high scores (0.636 and 0.673), while "Moderate" and "Neutral" posts show progressively lower scores, reflecting the model’s accuracy in quantifying intensity. The diverse subreddit sources demonstrate thematic variation, ensuring the model's ability to generalize across contexts. Moreover, emotionally charged language (e.g., ``genocide," ``pure evil") correlates strongly with higher extremism scores, indicating sensitivity to linguistic intensity. These aspects validate the dataset's design for testing the model's classification capability.

The statistics in Table \ref{tab:valStat} reveal how the components \(a\) (anger score), \(|p|\) (absolute polarity), and \(s\) (subjectivity) contribute to the extremism metrics. The mean values of \(a\) show an inconsistent relationship with post labels, as it is higher for Moderate than Extreme posts, which may indicate that anger alone is insufficient to distinguish these categories. However, \(|p|\) and \(s\) more effectively differentiate between labels, with Extreme posts showing the highest values for both. The normalization formula \(\chi_{\text{norm}}\) amplifies these differences, with a significant gap between Extreme (\(5.192\)) and Neutral (\(0.585\)) posts. Finally, \(\chi^l_{\text{norm}}\) refines this measure by incorporating the logarithm of word count (\(L\)), further distinguishing the categories and ensuring robustness across varying post lengths. This combined formula shows a balanced and effective method for capturing extremism, overcoming the limitations of individual components like anger.

\begin{table}[htbp]
  \centering
  \caption{Statistics for the extremism score and its components for each label in validation data.}
    \begin{tabular}{p{2cm}p{2.5cm}p{2.5cm}p{2.5cm}p{2.5cm}p{2.5cm}}
    \toprule
    \textbf{Label} & $a$ & $|p|$ & $s$ & $\chi_{\text{norm}}$ & $\chi^l_{\text{norm}}$ \\
    \toprule
    \textbf{Extreme} & 0.506 & 0.283 & 0.434 & 5.192 & 0.931 \\
    \textbf{Moderate} & 0.629 & 0.132 & 0.354 & 2.084 & 0.381 \\
    \textbf{Neutral} & 0.203 & 0.099 & 0.225 & 0.585 & 0.108 \\
    \bottomrule
    \end{tabular}%
  \label{tab:valStat}%
\end{table}%


\subsection{Interest in the Conflict on Reddit}
\begin{figure}[ht!]
\centering
\includegraphics[width=\linewidth]{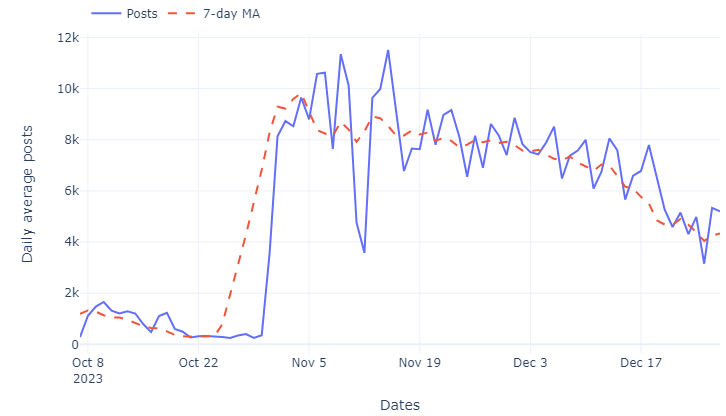}
\caption{Daily number of posts on Reddit subreddits related to the Israeli-Palestinian conflict. Dots represent daily counts, the solid line shows the 7-day moving average (MA), and the dashed line indicates the running mean. The period from October 30th to November 1st is highlighted by a red vertical rectangle.}
\label{fig:interest}
\end{figure}

Building on the insights gained from the statistical analysis of the extremism scores, we now turn our attention to examining how interest in the conflict evolved over time, as reflected in the daily posting activity on Reddit. The level of interest in the Israeli-Palestinian conflict on Reddit, as measured by the daily number of posts across four selected subreddits (\textit{r/IsraelPalestine}, \textit{r/worldnews}, \textit{r/Palestine}, and \textit{r/Judaism}), showed significant fluctuations over the observed period. Figure \ref{fig:interest} illustrates this dynamic, with daily post counts represented as dots, the 7-day moving average (MA) as a solid line, and the running mean as a dashed line.

Initially, from October 7th until the end of October, the daily post count remained relatively steady, with fewer than 2,000 posts per day. This indicates a period of low to moderate interest in the conflict within the Reddit community. However, a dramatic surge in the number of posts is observed starting from October 30th and 31st, coinciding with the IDF's bombings of Al Quds Hospital \citep{tbs1} and the Jabalia refugee camp \citep{october31} on these dates. The period between October 30th and November 1st, highlighted by a red vertical rectangle in Figure \ref{fig:interest}, marks the critical point where public interest peaked due to the intensification of the conflict. 

The number of daily posts jumped to above and around 10,000, reflecting a substantial spike in user engagement and discourse related to the conflict. The heightened level of interest continued through November, with the 7-day moving average remaining over 8,000 posts until December. This sustained interest underscores the significant impact of the aforementioned events and beyond on public attention and discussion within these subreddits. As the period progressed into December, the running mean levelled off to around 6,000 posts, with daily post counts stabilising between 4,000 and 5,000.

\subsection{Analysing Extreme opinions}
\begin{figure}[ht!]
\centering
\includegraphics[width=\linewidth]{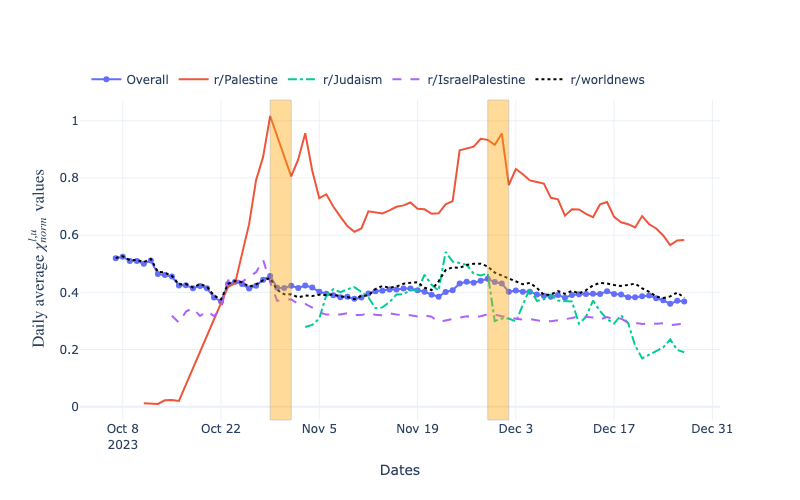}
\caption{7-day MA graphs of daily average extremism score for all four subreddits over time.}
\label{fig:extreme}
\end{figure}

In this experimental case, the proposed extremism score, $\chi_{norm}^{l,u}$, is examined over time for all four subreddits and in aggregate. Figure \ref{fig:extreme} provides a detailed analysis, showing the 7-day moving average (MA) of daily average $\chi_{norm}^{l,u}$ values for each subreddit and the overall dataset throughout the study period. The overall trend and the trend for \textit{r/worldnews} remain relatively stable over time, with $\chi_{norm}^{l,u}$ values consistently around 0.4-0.5, which is higher than the 75th percentile of the $\chi_{norm}^{l,u}$ values.

More interesting characteristics can be observed in the remaining three subreddits, which exhibit more biased tendencies compared to \textit{r/worldnews}. The trend for the \textit{r/IsraelPalestine} subreddit starts at a similar level to the overall trend, with a noticeable increase October 22nd where it reaches a peak towards the end of October and the beginning of November, coinciding with the onset of IDF's aggressive bombings. Starting from early-mid Noveber, the $\chi_{norm}^{l,u}$ value reaches approximately 0.3 and maintains a steady trend for the remainder of the analysis period.

When examining the Israeli-biased subreddit \textit{r/Judaism}, a more variable characteristic is observed compared to the previously discussed subreddits. The trend for \textit{r/Judaism} fluctuates, ranging from around 0.4 and occasionally dropping down to 0.2. This variability may indicate that the most extreme posts were not predominantly from this subreddit, or it could be due to the relatively low volume of posts, which might have obscured the display of this effect in our analysis.

Lastly, when examining the \textit{r/Palestine} subreddit, we observe radically different characteristics compared to the trends of all other subreddits. During the initial phases of the conflict, which overlap with the study period, this subreddit generally features non-extreme opinions with $\chi_{norm}^{l,u}$ values close to 0.0 until mid-October. However, starting with the increased aggression from the IDF close to the end of October, the extremism trend of this subreddit escalates sharply, reaching a $\chi_{norm}^{l,u}$ value around 1.0 within just two week, peaking on October 29. Although it stabilises after a drop to 0.7 until the end of November, another peak occurs on December 1, with $\chi_{norm}^{l,u}$ reaching values around 0.9. After this date, until the end of the study period, it follow a steady drop until values around 0.6. 

In addition to the aggression period of the IDF, highlighted by an orange vertical rectangle at the end of October, a second period is marked with another orange rectangle in late November and early December. This second interval coincides with the end of the ceasefire following a terrorist attack in West Jerusalem on November 30 \citep{nov30}. It is followed by renewed aggression from the IDF, including several ground operations and the bombing of the Jabalia refugee camp for the second time on December 3 \citep{dec3}.
 
To sum up, our analysis of extremism scores across different subreddits related to the Israeli-Palestinian conflict reveals significant trends that correspond closely with real-life events. The relatively stable extremism scores in the \textit{r/worldnews} subreddit contrast with the more variable and often higher scores in subreddits like \textit{r/IsraelPalestine} and \textit{r/Judaism}, reflecting the inherent biases within these communities. Notably, the \textit{r/Palestine} subreddit shows a dramatic increase in extremism scores following periods of intensified conflict, such as the IDF's bombings at the end of October and the renewed aggression in early December. These peaks in extremism scores validate our methodology, demonstrating a logical overlap between heightened conflict-related events and increased extreme opinions in social media discussions.

\subsection{IDF vs. Hamas in Sentiments}
\begin{figure}[ht!]
\centering
\includegraphics[width=\linewidth]{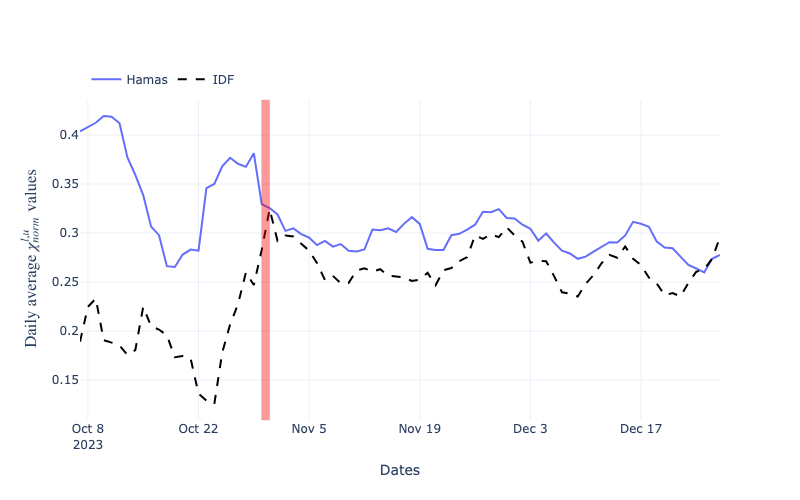}
\caption{7-day MA graphs of daily average extremism score for IDF and Hamas over time.}
\label{fig:hamasidf}
\end{figure}

For this experimental case, we constructed two sample sets from the overall data encompassing all posts from the four subreddits, focusing on mentions of IDF and Hamas. Posts containing the keyword ``Hamas'' were included in the Hamas sample set, while those with the keyword ``IDF'' formed the IDF sample set. We plotted their daily average 7-day moving average (MA) trends for the extremism score, $\chi_{norm}^{l,u}$, in Figure \ref{fig:hamasidf}. The period at the end of October is highlighted with a red vertical rectangle, marking a significant change point for both trends.

Analysis of the trends in Figure \ref{fig:hamasidf} reveals two distinct phases before and after the aggressive period at the end of October. Initially, the trends follow similar characteristics where $\chi_{norm}^{l,u}$ scores for Hamas oscillates between 0.3 and 0.4 whilst IDF related $\chi_{norm}^{l,u}$ scores halved but still oscillates between 0.1 and 0.2. However, the aggression commencing at the end of October alters this pattern, leading to synchronized trends where both experience similar increases and decreases. This shift suggests that escalating aggression and subsequent events triggered parallel fluctuations in extreme opinions within both communities.


\subsection{Quantitative Effects of Each Score on Extremism Score}
\begin{figure}[ht!]
\centering
\includegraphics[width=\linewidth]{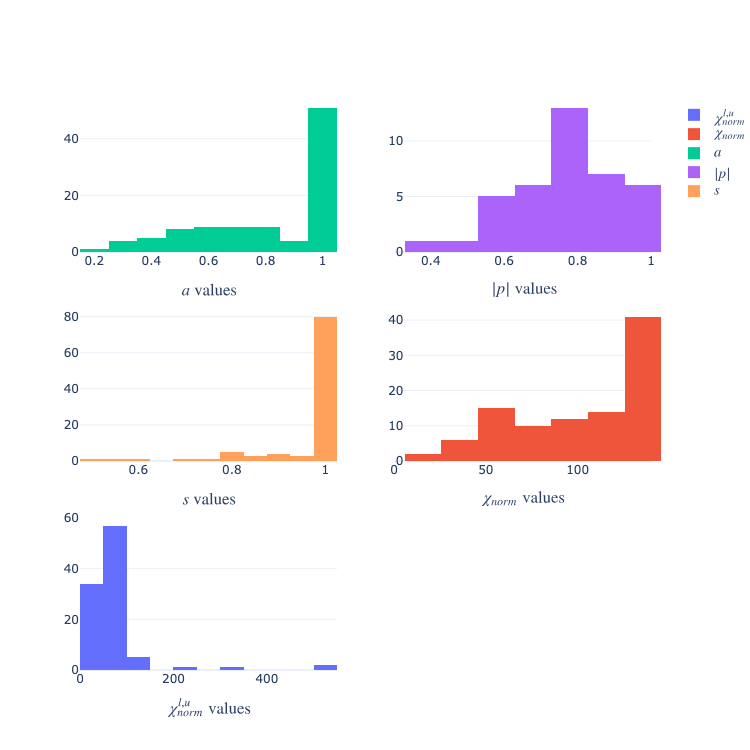}
\caption{The score distributions of the top 100 posts with the highest final extremism score.}
\label{fig:dists}
\end{figure}

In this subsection, we analyse the quantitative effects of each score—$a$, $|p|$, $(1 - s)$, and $\chi_{norm}$—on the weighted normalised extremism score of $\chi_{norm}^{l,u}$. This analysis is based on Figure \ref{fig:dists} consisting of five histograms, each representing the distribution of these scores for the top 100 posts ranked by $\chi_{norm}^{l,u}$.

Examining the left top sub-figure in Figure \ref{fig:dists}, we can note that the histogram for $a$ values shows a prominent peak in the range of 0.95+, encompassing 51 posts. All remaining posts are distributed uniformly especially between 0.4-0.8 with 40 posts. This distribution indicates that higher $a$ scores are more prevalent among the top posts, suggesting a significant correlation between high $a$ levels and extreme opinions.

Looking at the top right sub-figure in Figure \ref{fig:dists}, we see that the absolute polarity, $|p|$, histogram reveals a single notable peak where the number of posts exceeds 10. This peak occurs in the range 0.70-0.799, with 13 posts. The remaining $|p|$ values have fewer than 10 posts each mostly distributed for $|p|$ values higher than 0.5. This distribution indicates that moderate $|p|$ values are more common in posts with high extremism scores, highlighting the role of balanced sentiment in contributing to extreme opinions. The histogram for $s$ (middle-left in Figure \ref{fig:dists}) displays a right-skewed characteristics with a single notable peak with 80 posts in the range 0.975-1.0. This suggests that posts with high levels of subjectivity are more likely to have high extremism scores.

The normalised extremism, $\chi_{norm}$, histogram plotted in the middle-right in Figure \ref{fig:dists} is right-skewed, with higher values having a higher number of posts. Specifically, $\chi_{norm}$ values higher than 40.0 accommodates more than 90 posts at the range 120.0-139.9 with 41 posts, and at 40.0-120.0 with 51 posts. This skewness indicates that most posts with high extremism scores not only have similarly higher $\chi_{norm}$ values but also emphasise that even moderate $\chi_{norm}$-valued posts are significantly represented in the top-ranked set.

To sum up the all findings from Figure \ref{fig:dists}, our analysis reveals that higher $a$ and $s$ scores, and moderate $|p|$ values are strongly associated with high $\chi_{norm}^{l,u}$ scores. Additionally, the right-skewed distribution of $\chi_{norm}$ indicates that even moderately extreme posts play a significant role in the top-ranked set, providing valuable insights into the characteristics and interrelationships of these scores.

\subsection{Corpus Similarity}
For this experimental case, we created individual corpora for the top 1000 posts of each score. After several text processing operations, including the removal of punctuation, stopwords, and unnecessary spaces using Python's \textit{string} and \textit{regex} modules, we analysed these corpora to identify frequently used words and calculate the Jaccard index between each pair of corpora.

\begin{figure}[ht!]
\centering
\subfloat[$a$]{\includegraphics[width=0.49\linewidth]{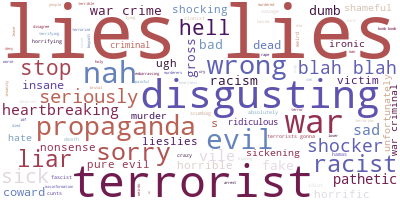}\label{fig:WC1}}
\subfloat[$|p|$]{\includegraphics[width=0.49\linewidth]{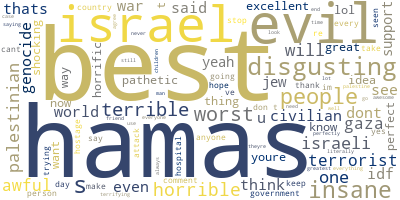}\label{fig:WC2}}\\
\subfloat[$s$]{\includegraphics[width=0.49\linewidth]{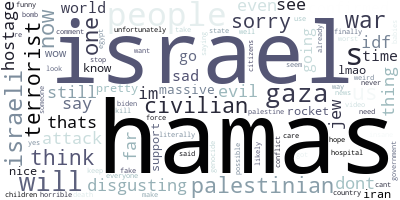}\label{fig:WC3}}
\subfloat[$\chi_{norm}$]{\includegraphics[width=0.49\linewidth]{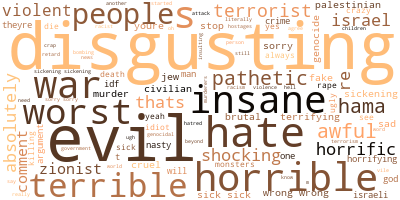}\label{fig:WC4}}\\
\subfloat[$\chi_{norm}^{l,u}$]{\includegraphics[width=0.49\linewidth]{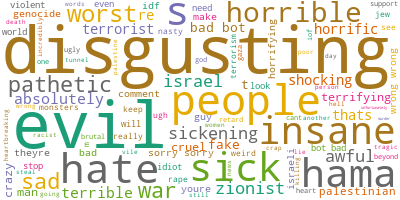}\label{fig:WC5}}
\caption{Wordcloud distributions of each score for the top 1000 highest posts.}\label{fig:WCall}
\end{figure}

The word clouds for each score's corpus are depicted in Figure \ref{fig:WCall}, providing a visual representation of the most commonly used words. Additionally, we calculated the Jaccard index to investigate the similarities between the corpora, which is detailed in a heatmap representation below in Figure \ref{fig:jaccard}. The Jaccard index, while rarely used in text processing, was intentionally selected to measure the percentage of word overlap between different score corpora.

\begin{figure}[ht!]
\centering
\includegraphics[width=.75\linewidth]{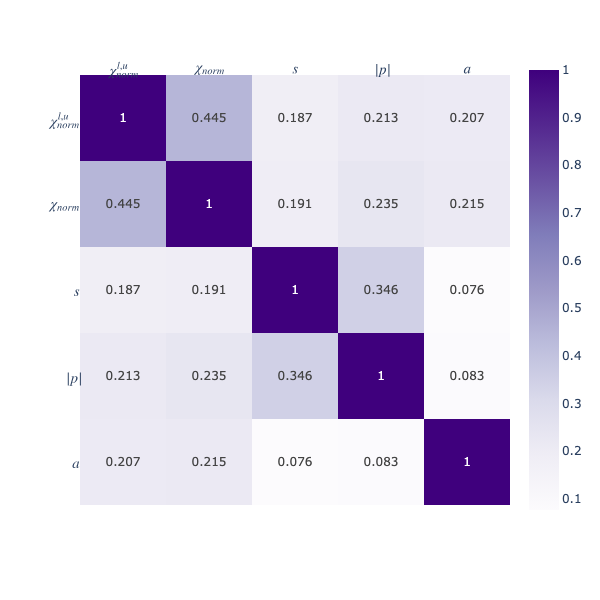}
\caption{Jaccard Similarity score heatmap representation for each corpus obtained from different scores.}
\label{fig:jaccard}
\end{figure}

From the most frequently used words in each score's corpus, we observe notable differences and similarities. For instance, terms like 'disgusting', 'war', 'terrorist', and 'people' appear across multiple corpora, indicating common themes of conflict and aggression. However, unique words in specific corpora, such as 'lies lies' in anger or 'evil' in absolute polarity, highlight distinct emotional and sentiment-driven perspectives. The Jaccard index matrix provides a quantitative measure of the overlap between different corpora. The highest similarity is observed between the normalised extremism and weighted normalised extremism scores (0.445), indicating that posts with high extremism scores, whether weighted or not, share a significant proportion of common words. The anger corpus exhibits the least overlap with the \(|p|\) (polarity) and \(s\) (subjectivity) corpora, as reflected by relatively low Jaccard index values. In contrast, \(|p|\) and \(s\) share a considerable overlap, with a Jaccard index of 0.346. This suggests that posts characterized by high anger levels tend to employ a more distinct vocabulary, differing significantly from the lexicon associated with polarity and subjectivity measures.

These findings underscore the complexity and multifaceted nature of extreme opinions. While there are common threads in the discourse around the Israeli-Palestinian conflict, the specific emotional and sentiment nuances captured by different scores reveal a rich diversity in the language used. This analysis provides a deeper understanding of how various factors contribute to the expression of extremism in social media posts.

\subsection{Chronological Validation}
\begin{figure}[htbp!]
\centering
\includegraphics[width=\linewidth]{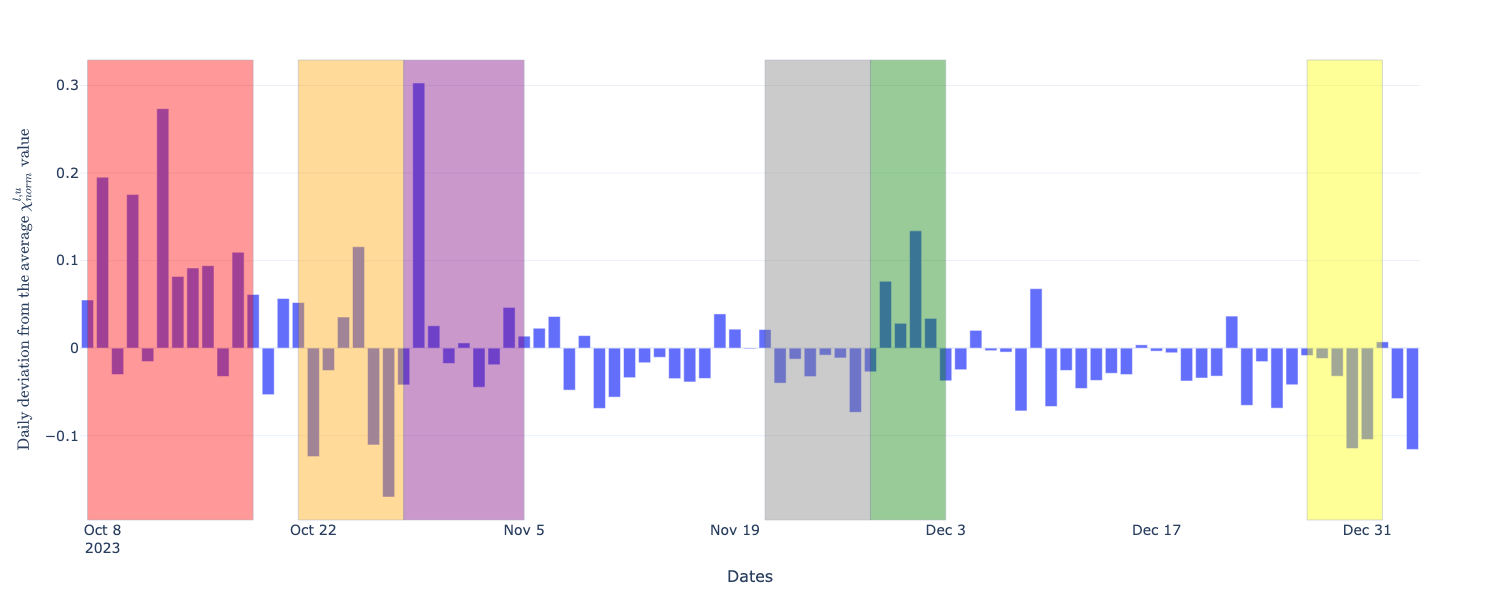}
\caption{Daily deviation from the average $\chi_{norm}^{l,u}$ value.}
\label{fig:chrono}
\end{figure}

We previously highlighted two important breaking points during the study period: (1) the IDF's aggression and bombings of a hospital and refugee camps in late October, and (2) the ending of the ceasefire following a terrorist attack in Israel and the subsequent increase in IDF aggression during late November and early December. In this final experimental analysis case, we examine the daily fluctuations above and below average scores depicted in Figure \ref{fig:chrono} and investigate additional real-life events that correspond with these increases and decreases in the proposed extremism scores. Additionally, Table \ref{tab:appendix} provides detailed chronological events within the study timeframe with newspaper references. Please refer to Table \ref{tab:appendix} for daily event details. 
 
In the early stages of the conflict, Figure \ref{fig:chrono} shows four significant increases (0.1+ - red region) in the $\chi_{norm}^{l,u}$ score on the 8th, 10th, 12th, and 17th of October 2023. During this period, Israel began to take action following the Hamas surprise attack on the 7th of October. The increase on the 8th is likely related to the Hamas surprise attack. The increases on the 10th, 12th and 17th seem to be reactions to (i) Israel's aggressive actions against Gaza City, such as cutting off electricity and gas \citep{bbc3}, and the explosion at the Al Ahli Hospital, which killed 500 people and was used as a shelter for thousands of displaced residents \citep{october17}, and (ii) the aftermath and death toll of the Hamas attack on a music festival in South Israel that started the conflict \citep{october9}.

The subsequent period, particularly until the end of October, mostly involved global humanitarian appeals, with a few hostage releases accompanying these efforts. These positive steps likely created hope within the community, resulting in several significant drops in the extremism score (to -0.12 and -0.17 - orange region). However, this trend was abruptly reversed when Israel decided to expand their ground operations in Gaza, followed by bombings at Al Quds Hospital and the Jabalia Refugee Camp, which likely caused several increases in the extremism scores (purple region).

From this period up to late November, fluctuations were relatively minor. In November, two key periods need to be highlighted: (i) the ceasefire period, and (ii) the end of the ceasefire following a terrorist attack in Israel. As expected, the ceasefire period resulted in several consecutive negative extremism bars (grey region). Unfortunately, a terrorist attack in Israel on the 30th of November brought an abrupt end to the ceasefire, which also led to renewed Israeli aggression at the end of November and in the first few days of December. In our analysis, this period is shown within a green region, consisting of several consecutive positive values.

The final period we highlight encompasses the last few days of the year. This period is dominated by increasing public support for Palestine and the efforts of organisations and countries like MSF (Médecins Sans Frontières/Doctors Without Borders) \citep{dec8}, the UN \citep{dec28}, The Red Crescent \citep{aj1}, and South Africa \citep{reuters2}, which raised awareness about the humanitarian crisis in Gaza. This support, despite continued minor aggressions from both sides, is likely the reason for several consecutive drops in extremism values in Figure \ref{fig:chrono} (yellow region).

The analysis of daily extremism score fluctuations reveals a strong correlation between significant real-life events and the spikes and drops in the extremism scores. This chronological validation highlights the responsiveness of the proposed extremism measurement to real-world developments, reinforcing its potential as a robust tool for analysing the dynamics of extreme opinions in social media discussions.

\section{Critical Discussion and Conclusions}\label{sec:critical}
The proposed approach for investigating extreme opinions on social media posts, particularly in the context of the Israeli-Palestinian conflict, demonstrates several strengths. One of the key positives is the comprehensive and multi-faceted scoring system developed to quantify extremism. By incorporating various dimensions such as anger, absolute polarity, subjectivity, and normalised extremism, we were able to capture a nuanced picture of online discourse. The use of weighted normalisation further allowed for a refined analysis, adjusting for the relative importance of each factor.

Moreover, the temporal analysis provided valuable insights into how extremism scores fluctuated in response to significant real-world events. For instance, the dramatic spike in extremism scores following the IDF's bombings of Al Quds Hospital and Jabalia refugee camp highlighted the direct correlation between violent incidents and the escalation of extreme opinions online. This temporal mapping offers a robust framework for understanding the dynamics of online extremism in the context of real-world triggers.

The corpus-based analysis, including the construction of word clouds and the calculation of the Jaccard index, offered a deeper exploration of the language used in extreme posts. This approach not only identified common themes and unique vocabulary associated with different scores but also revealed the interplay between emotional and sentiment-driven perspectives in the expression of extremism.

Despite these strengths, the approach has several limitations and challenges. One major limitation is the reliance on keyword-based selection for creating sample sets, such as those for IDF and Hamas. This method may introduce bias, as it assumes that the presence of specific keywords is indicative of the post's context and sentiment, which may not always be accurate. Additionally, the use of a predefined list of swear words and the subsequent censoring or removal may overlook the evolving nature of language and the emergence of new terms used to express extreme opinions.

Another challenge is the inherent subjectivity in determining the weights for normalisation. While the weights were chosen based on logical reasoning and prior research, they may still introduce bias and affect the robustness of the extremism score. Furthermore, the analysis is constrained by the availability and quality of data. Posts from only four subreddits were considered, which may not be fully representative of the broader social media landscape. This limitation calls for a broader data collection to enhance the generalisability of the findings.

The approach also faces challenges in accurately capturing the sentiment and subjectivity of posts. Natural language processing tools, despite their advancements, still struggle with the nuances of human language, especially in the context of sarcasm, irony, and cultural references. This can lead to misclassification and affect the overall accuracy of the extremism score.

In summary, while the proposed approach offers a detailed and dynamic method for analysing extreme opinions on social media, it is not without its limitations. Future work should aim to address these challenges by incorporating more sophisticated natural language processing techniques, expanding the scope of data collection, and refining the weighting system for normalisation. Additionally, exploring the integration of machine learning models to automatically detect and adapt to new forms of extreme language could further enhance the robustness of the analysis. Despite the challenges, this approach provides a valuable foundation for understanding the complex landscape of online extremism and its relationship with real-world events.






\appendix
\section[\appendixname~\thesection]{Choronological details of events with references}
This section presents significant events in chronological order with newspaper references for the timeframe covered by this paper. Refer to Table \ref{tab:appendix} for more details.

\begin{longtable}{p{2.5cm}p{10cm}p{3cm}}
    \caption{Chronological Events in the Israeli-Palestinian Conflict} \label{tab:appendix}\\
    \toprule
    \textbf{Date} & \textbf{Event Details} & \textbf{Reference} \\
    \toprule
    \endfirsthead

    \toprule
    \textbf{Date} & \textbf{Event Details} & \textbf{Reference} \\
    \toprule
    \endhead

    \midrule
    \multicolumn{3}{r}{Continued on next page} \\
    \midrule
    \endfoot

    \bottomrule
    \endlastfoot

    October 7, 2023 & Surprise Hamas attack. & \citet{october7} \\\hline

    October 7, 2023 &  The leader of the Al-Qassam brigades release a statement taking responsibility for the attack & \citet{october7a} \\\hline

    October 7, 2023 & Benjamin Netanyahu declares that Israel is at war & \citet{october7b} \\\hline

    October 8, 2023 & Israel cut off electricity and gas from Gaza & \citet{october8} \\\hline

    October 8, 2023 & Netanyahu vows for ``Mighty vengeance" whilst at least 480 Israelis and Palestinians dead & \citet{october8a} \\\hline
    
    October 9, 2023 & 260 bodies were recovered from the Israeli music festival site. & \citet{october9} \\\hline
    
    October 10, 2023 & the World Health Organization asks for access to aid to the Gaza Strip. & \citet{october10} \\\hline
        
    October 12, 2023 & Human Rights Watch accuses Israel of using white phosphorus. & \citet{october12} \\\hline
    
    October 13, 2023 & IDF calls for Gaza citizens to evacuate north of the strip. & \citet{october13} \\\hline

    October 13, 2023 & Israel declares to expand the attack on Gaza with offensive operations through land and sea. & \citet{october13a} \\\hline

    October 14, 2023 & Israel attacked Khar Younis (the city in the south, after the day they told people to flee south). & \citet{october14} \\\hline

    October 15, 2023 & Iran warns Israel of regional escalation. & \citet{october15} \\\hline

    October 16, 2023 & the White House announced Biden would visit Israel. & \citet{october16} \\\hline

    October 17, 2023 & 500 people are killed in an explosion at the Al Ahli hospital in the middle of Gaza. & \citet{october17} \\\hline
    
    October 18, 2023 & Biden arrives in Israel and meets Netanyahu confirming support. & \citet{october18} \\\hline

    October 19, 2023 & An Orthodox church used as a shelter was bombed by IDF, dozens of deaths. & \citet{october19} \\\hline

    October 20, 2023 & Hamas releases 2 American hostages. & \citet{october20} \\\hline
    
    October 21, 2023 & Rafah crossing opens allowing for the entrance of 20 trucks. & \citet{october21} \\\hline
    
    October 22, 2023 & The UN says that 406K people are sheltering in UNRWA structures, also calls for more help and accuses Israel. & \citet{october22} \\\hline
    
    October 24, 2023 & UN Chief Guterres makes resonant declarations indirectly accusing Israel of war crimes. & \citet{october24} \\\hline
        
    October 25, 2023 & Israel bombs Southern Gaza after warning civilians to escape there.?. & \citet{october25} \\\hline
        
    October 27, 2023 & The second ground operation in Gaza for Israel. & \citet{october27} \\\hline
    
    October 28, 2023 & Major developments in the ongoing conflict are reported, with the military actions continuing to escalate. & \citet{october28} \\\hline
    
    October 29, 2023 & Palestine Red Crescent reports Israeli orders for the evacuation of Al-Quds Hospital. & \citet{october29} \\\hline
    
    October 30, 2023 & Heavy clashes as Israeli tanks reach Gaza City outskirts and cut a key road. & \citet{october30} \\\hline

    October 30, 2023 & Israeli air strikes in the area around Al-Quds Hospital, which Israel has told doctors to evacuate. & \citet{tbs1} \\\hline
    
    October 31, 2023 & Israel bombs the Jabalia refugee camp, claiming to kill a Hamas commander. & \citet{october31} \\\hline
    
    November 1, 2023 & Details emerge on Israel's deadly attack on the Jabalia refugee camp. & \citet{november1} \\\hline
    
    November 2, 2023 & Gaza reports Israeli strikes on refugee camp killed more than 195 people. & \citet{november2} \\\hline

    November 3, 2023 & Israel raids in West Bank refugee camps, at least 11 killed & \citet{nov3} \\\hline
    November 5, 2023 & Major humanitarian organisations call for an immediate ceasefire in Gaza & \citet{nov5} \\\hline
    November 7, 2023 & Netanyahu announces plans for Israel's security responsibility in Gaza post-war & \citet{nov7} \\\hline
    November 8, 2023 & IDF troops move into the heart of Gaza & \citet{nov8} \\\hline
    November 9, 2023 & Increase in violence by Israeli settlers in West Bank reported & \citet{nov9} \\\hline
    November 12, 2023 & UN agencies call for a stop to attacks on health care in Gaza & \citet{nov12} \\\hline
    November 15, 2023 & Israeli troops enter Gaza's biggest hospital, Al Shifa & \citet{nov15} \\\hline
    November 16, 2023 & Telecommunications stop working in Gaza & \citet{nov16} \\\hline
    November 17, 2023 & Airstrike on Al Falah school in Gaza, 20 dead & \citet{nov17} \\\hline
    November 20, 2023 & Airstrike on the last working hospital in north Gaza & \citet{nov20} \\\hline
    November 21, 2023 & Israel government agrees to hostage deal & \citet{nov21} \\\hline
    November 24, 2023 & Ceasefire over prisoners deal starts & \citet{nov24} \\\hline
    November 30, 2023 & Terrorist attack in West Jerusalem & \citet{nov30} \\\hline
    December 1, 2023 & IDF resumes strikes & \citet{dec1} \\\hline
    December 2, 2023 & Israel withdraws from talks in Qatar & \citet{dec2} \\\hline
    December 3, 2023 & IDF attacks Jabalia refugee camp & \citet{dec3} \\\hline
    December 4, 2023 & IDF demolishes the Palestinian Palace of Justice & \citet{dec4} \\\hline
    December 5, 2023 & Another UNRWA school bombed & \citet{dec5} \\\hline
    December 6, 2023 & An Israeli attack in the Jabalia refugee camp kills 22 members of Al Jazeera correspondent & \citet{dec6} \\\hline
    December 7, 2023 & The great mosque of Gaza is destroyed under IDF bombing & \citet{dec7} \\\hline
    December 8, 2023 & The USA vetoes a UN resolution asking for a ceasefire & \citet{dec8} \\\hline
    December 11, 2023 & Health system collapsing & \citet{dec11} \\\hline
    December 12, 2023 & Joint statement from Canada, Australia, and New Zealand asking for a ceasefire & \citet{dec12} \\\hline
    December 17, 2023 & IDF finds a tunnel large enough for vehicles near the border & \citet{dec17} \\\hline
    December 19, 2023 & Other Israeli attacks & \citet{dec19} \\\hline
    December 20, 2023 & Hamas asks for a ceasefire in exchange for hostages, Israel refuses & \citet{dec20} \\\hline
    December 28, 2023 & UN report warns of rapidly deteriorating human rights situation in the West Bank & \citet{dec28} \\
\end{longtable}

\bibliographystyle{elsarticle-num-names}
\bibliography{refs.bib}

\begin{thebibliography}{114}
\expandafter\ifx\csname natexlab\endcsname\relax\def\natexlab#1{#1}\fi
\providecommand{\url}[1]{\texttt{#1}}
\providecommand{\href}[2]{#2}
\providecommand{\path}[1]{#1}
\providecommand{\DOIprefix}{doi:}
\providecommand{\ArXivprefix}{arXiv:}
\providecommand{\URLprefix}{URL: }
\providecommand{\Pubmedprefix}{pmid:}
\providecommand{\doi}[1]{\href{http://dx.doi.org/#1}{\path{#1}}}
\providecommand{\Pubmed}[1]{\href{pmid:#1}{\path{#1}}}
\providecommand{\bibinfo}[2]{#2}
\ifx\xfnm\relax \def\xfnm[#1]{\unskip,\space#1}\fi
\bibitem[{Liyih et~al.(2024)Liyih, Anagaw, Yibeyin, and Tehone}]{liyih2024sentiment}
\bibinfo{author}{A.~Liyih}, \bibinfo{author}{S.~Anagaw}, \bibinfo{author}{M.~Yibeyin}, \bibinfo{author}{Y.~Tehone},
\newblock \bibinfo{title}{{Sentiment analysis of the Hamas-Israel war on YouTube comments using deep learning}},
\newblock \bibinfo{journal}{Scientific Reports} \bibinfo{volume}{14} (\bibinfo{year}{2024}) \bibinfo{pages}{13647}.
\bibitem[{Chen et~al.(2024)Chen, He, Burghardt, Zhang, and Lerman}]{chen2024isamasred}
\bibinfo{author}{K.~Chen}, \bibinfo{author}{Z.~He}, \bibinfo{author}{K.~Burghardt}, \bibinfo{author}{J.~Zhang}, \bibinfo{author}{K.~Lerman},
\newblock \bibinfo{title}{{IsamasRed: A Public Dataset Tracking Reddit Discussions on Israel-Hamas Conflict}},
\newblock in: \bibinfo{booktitle}{Proceedings of the International AAAI Conference on Web and Social Media}, volume~\bibinfo{volume}{18}, \bibinfo{year}{2024}, pp. \bibinfo{pages}{1900--1912}.
\bibitem[{Tschirky and Makhortykh(2024)}]{tschirky2024azovsteel}
\bibinfo{author}{M.~Tschirky}, \bibinfo{author}{M.~Makhortykh},
\newblock \bibinfo{title}{{\# Azovsteel: Comparing qualitative and quantitative approaches for studying framing of the siege of Mariupol on Twitter}},
\newblock \bibinfo{journal}{Media, War \& Conflict} \bibinfo{volume}{17} (\bibinfo{year}{2024}) \bibinfo{pages}{163--178}.
\bibitem[{Aviv and Ferri(2023)}]{aviv2023russian}
\bibinfo{author}{I.~Aviv}, \bibinfo{author}{U.~Ferri},
\newblock \bibinfo{title}{{Russian-Ukraine armed conflict: Lessons learned on the digital ecosystem}},
\newblock \bibinfo{journal}{International Journal of Critical Infrastructure Protection} \bibinfo{volume}{43} (\bibinfo{year}{2023}) \bibinfo{pages}{100637}.
\bibitem[{Schoemaker and Stremlau(2014)}]{schoemaker2014media}
\bibinfo{author}{E.~Schoemaker}, \bibinfo{author}{N.~Stremlau},
\newblock \bibinfo{title}{Media and conflict: An assessment of the evidence},
\newblock \bibinfo{journal}{Progress in Development Studies} \bibinfo{volume}{14} (\bibinfo{year}{2014}) \bibinfo{pages}{181--195}.
\bibitem[{Hawkins(2002)}]{hawkins2002other}
\bibinfo{author}{V.~Hawkins},
\newblock \bibinfo{title}{The other side of the cnn factor: the media and conflict},
\newblock \bibinfo{journal}{Journalism Studies} \bibinfo{volume}{3} (\bibinfo{year}{2002}) \bibinfo{pages}{225--240}.
\bibitem[{Zeitzoff(2017)}]{zeitzoff2017social}
\bibinfo{author}{T.~Zeitzoff},
\newblock \bibinfo{title}{How social media is changing conflict},
\newblock \bibinfo{journal}{Journal of Conflict Resolution} \bibinfo{volume}{61} (\bibinfo{year}{2017}) \bibinfo{pages}{1970--1991}.
\bibitem[{Vera et~al.(2018)Vera, Su{\'a}rez, and Lopera}]{vera2018sentiment}
\bibinfo{author}{V.~D.~G. Vera}, \bibinfo{author}{L.~M.~M. Su{\'a}rez}, \bibinfo{author}{I.~C.~P. Lopera},
\newblock \bibinfo{title}{{Sentiment analysis on post conflict in Colombia: A text mining approach}},
\newblock \bibinfo{journal}{Asian Journal of Applied Sciences} \bibinfo{volume}{6} (\bibinfo{year}{2018}).
\bibitem[{{\"O}zt{\"u}rk and Ayvaz(2018)}]{ozturk2018sentiment}
\bibinfo{author}{N.~{\"O}zt{\"u}rk}, \bibinfo{author}{S.~Ayvaz},
\newblock \bibinfo{title}{{Sentiment analysis on Twitter: A text mining approach to the Syrian refugee crisis}},
\newblock \bibinfo{journal}{Telematics and Informatics} \bibinfo{volume}{35} (\bibinfo{year}{2018}) \bibinfo{pages}{136--147}.
\bibitem[{Aslan(2023)}]{aslan2023deep}
\bibinfo{author}{S.~Aslan},
\newblock \bibinfo{title}{{A deep learning-based sentiment analysis approach (MF-CNN-BILSTM) and topic modeling of tweets related to the Ukraine--Russia conflict}},
\newblock \bibinfo{journal}{Applied Soft Computing} \bibinfo{volume}{143} (\bibinfo{year}{2023}) \bibinfo{pages}{110404}.
\bibitem[{Akpatsa et~al.(2022)Akpatsa, Addo, Lei, Li, Dorgbefu~Jr, Fiawoo, Nartey, and Dagadu}]{akpatsa2022sentiment}
\bibinfo{author}{S.~K. Akpatsa}, \bibinfo{author}{P.~C. Addo}, \bibinfo{author}{H.~Lei}, \bibinfo{author}{X.~Li}, \bibinfo{author}{M.~Dorgbefu~Jr}, \bibinfo{author}{D.~D. Fiawoo}, \bibinfo{author}{J.~Nartey}, \bibinfo{author}{J.~C. Dagadu},
\newblock \bibinfo{title}{{Sentiment Analysis and Topic Modeling of Twitter Data: A Text Mining Approach to the US-Afghan War Crisis}},
\newblock \bibinfo{journal}{Available at SSRN 4064560}  (\bibinfo{year}{2022}).
\bibitem[{Guerra and Karaku{\c{s}}(2023)}]{guerra2023sentiment}
\bibinfo{author}{A.~Guerra}, \bibinfo{author}{O.~Karaku{\c{s}}},
\newblock \bibinfo{title}{{Sentiment analysis for measuring hope and fear from Reddit posts during the 2022 Russo-Ukrainian conflict}},
\newblock \bibinfo{journal}{Frontiers in Artificial Intelligence} \bibinfo{volume}{6} (\bibinfo{year}{2023}) \bibinfo{pages}{1163577}.
\bibitem[{Del~Vicario et~al.(2016)Del~Vicario, Bessi, Zollo, Petroni, Scala, Caldarelli, Stanley, and Quattrociocchi}]{del2016spreading}
\bibinfo{author}{M.~Del~Vicario}, \bibinfo{author}{A.~Bessi}, \bibinfo{author}{F.~Zollo}, \bibinfo{author}{F.~Petroni}, \bibinfo{author}{A.~Scala}, \bibinfo{author}{G.~Caldarelli}, \bibinfo{author}{H.~E. Stanley}, \bibinfo{author}{W.~Quattrociocchi},
\newblock \bibinfo{title}{The spreading of misinformation online},
\newblock \bibinfo{journal}{Proceedings of the National Academy of Sciences} \bibinfo{volume}{113} (\bibinfo{year}{2016}) \bibinfo{pages}{554--559}.
\bibitem[{Matakos et~al.(2017)Matakos, Terzi, and Tsaparas}]{matakos2017measuring}
\bibinfo{author}{A.~Matakos}, \bibinfo{author}{E.~Terzi}, \bibinfo{author}{P.~Tsaparas},
\newblock \bibinfo{title}{Measuring and moderating opinion polarization in social networks},
\newblock \bibinfo{journal}{Data Mining and Knowledge Discovery} \bibinfo{volume}{31} (\bibinfo{year}{2017}) \bibinfo{pages}{1480--1505}.
\bibitem[{Medhat et~al.(2014)Medhat, Hassan, and Korashy}]{MEDHAT20141093}
\bibinfo{author}{W.~Medhat}, \bibinfo{author}{A.~Hassan}, \bibinfo{author}{H.~Korashy},
\newblock \bibinfo{title}{Sentiment analysis algorithms and applications: A survey},
\newblock \bibinfo{journal}{Ain Shams Engineering Journal} \bibinfo{volume}{5} (\bibinfo{year}{2014}) \bibinfo{pages}{1093--1113}. \URLprefix \url{https://www.sciencedirect.com/science/article/pii/S2090447914000550}. \DOIprefix\doi{https://doi.org/10.1016/j.asej.2014.04.011}.
\bibitem[{Liu(2020)}]{liu2020sentiment}
\bibinfo{author}{B.~Liu}, \bibinfo{title}{Sentiment analysis: Mining opinions, sentiments, and emotions}, \bibinfo{publisher}{Cambridge university press}, \bibinfo{year}{2020}.
\bibitem[{Nasukawa and Yi(2003)}]{nasukawa2003sentiment}
\bibinfo{author}{T.~Nasukawa}, \bibinfo{author}{J.~Yi},
\newblock \bibinfo{title}{Sentiment analysis: Capturing favorability using natural language processing},
\newblock in: \bibinfo{booktitle}{Proceedings of the 2nd international conference on Knowledge capture}, \bibinfo{year}{2003}, pp. \bibinfo{pages}{70--77}.
\bibitem[{Dave et~al.(2003)Dave, Lawrence, and Pennock}]{dave2003mining}
\bibinfo{author}{K.~Dave}, \bibinfo{author}{S.~Lawrence}, \bibinfo{author}{D.~M. Pennock},
\newblock \bibinfo{title}{Mining the peanut gallery: Opinion extraction and semantic classification of product reviews},
\newblock in: \bibinfo{booktitle}{Proceedings of the 12th international conference on World Wide Web}, \bibinfo{year}{2003}, pp. \bibinfo{pages}{519--528}.
\bibitem[{Hearst(1999)}]{hearst1999untangling}
\bibinfo{author}{M.~A. Hearst},
\newblock \bibinfo{title}{Untangling text data mining},
\newblock in: \bibinfo{booktitle}{Proceedings of the 37th Annual meeting of the Association for Computational Linguistics}, \bibinfo{year}{1999}, pp. \bibinfo{pages}{3--10}.
\bibitem[{Rehurek and Sojka(2010)}]{rehurek2010software}
\bibinfo{author}{R.~Rehurek}, \bibinfo{author}{P.~Sojka},
\newblock \bibinfo{title}{Software framework for topic modelling with large corpora},
\newblock in: \bibinfo{booktitle}{In Proceedings of the LREC 2010 workshop on new challenges for NLP frameworks}, \bibinfo{organization}{Citeseer}, \bibinfo{year}{2010}, pp. \bibinfo{pages}{46--50}.
\bibitem[{Feldman(2013)}]{feldman2013techniques}
\bibinfo{author}{R.~Feldman},
\newblock \bibinfo{title}{Techniques and applications for sentiment analysis},
\newblock \bibinfo{journal}{Communications of the ACM} \bibinfo{volume}{56} (\bibinfo{year}{2013}) \bibinfo{pages}{82--89}.
\bibitem[{Zucco et~al.(2017)Zucco, Calabrese, and Cannataro}]{zucco2017sentiment}
\bibinfo{author}{C.~Zucco}, \bibinfo{author}{B.~Calabrese}, \bibinfo{author}{M.~Cannataro},
\newblock \bibinfo{title}{Sentiment analysis and affective computing for depression monitoring},
\newblock in: \bibinfo{booktitle}{2017 IEEE international conference on bioinformatics and biomedicine (BIBM)}, \bibinfo{organization}{IEEE}, \bibinfo{year}{2017}, pp. \bibinfo{pages}{1988--1995}.
\bibitem[{Yu and Wang(2015)}]{yu2015world}
\bibinfo{author}{Y.~Yu}, \bibinfo{author}{X.~Wang},
\newblock \bibinfo{title}{World cup 2014 in the twitter world: A big data analysis of sentiments in us sports fans’ tweets},
\newblock \bibinfo{journal}{Computers in Human Behavior} \bibinfo{volume}{48} (\bibinfo{year}{2015}) \bibinfo{pages}{392--400}.
\bibitem[{Peng et~al.(2021)Peng, Cao, Zhou, Ouyang, Yang, Li, Jia, and Yu}]{peng2021survey}
\bibinfo{author}{S.~Peng}, \bibinfo{author}{L.~Cao}, \bibinfo{author}{Y.~Zhou}, \bibinfo{author}{Z.~Ouyang}, \bibinfo{author}{A.~Yang}, \bibinfo{author}{X.~Li}, \bibinfo{author}{W.~Jia}, \bibinfo{author}{S.~Yu},
\newblock \bibinfo{title}{A survey on deep learning for textual emotion analysis in social networks},
\newblock \bibinfo{journal}{Digital Communications and Networks}  (\bibinfo{year}{2021}).
\bibitem[{Thelwall et~al.(2010)Thelwall, Wilkinson, and Uppal}]{thelwall2010data}
\bibinfo{author}{M.~Thelwall}, \bibinfo{author}{D.~Wilkinson}, \bibinfo{author}{S.~Uppal},
\newblock \bibinfo{title}{Data mining emotion in social network communication: Gender differences in myspace},
\newblock \bibinfo{journal}{Journal of the American Society for Information Science and Technology} \bibinfo{volume}{61} (\bibinfo{year}{2010}) \bibinfo{pages}{190--199}.
\bibitem[{Pagolu et~al.(2016)Pagolu, Reddy, Panda, and Majhi}]{pagolu2016sentiment}
\bibinfo{author}{V.~S. Pagolu}, \bibinfo{author}{K.~N. Reddy}, \bibinfo{author}{G.~Panda}, \bibinfo{author}{B.~Majhi},
\newblock \bibinfo{title}{Sentiment analysis of twitter data for predicting stock market movements},
\newblock in: \bibinfo{booktitle}{2016 international conference on signal processing, communication, power and embedded system (SCOPES)}, \bibinfo{organization}{IEEE}, \bibinfo{year}{2016}, pp. \bibinfo{pages}{1345--1350}.
\bibitem[{Ji and Han(2022)}]{10.3389/frai.2022.884699}
\bibinfo{author}{R.~Ji}, \bibinfo{author}{Q.~Han},
\newblock \bibinfo{title}{Understanding heterogeneity of investor sentiment on social media: A structural topic modeling approach},
\newblock \bibinfo{journal}{Frontiers in Artificial Intelligence} \bibinfo{volume}{5} (\bibinfo{year}{2022}). \URLprefix \url{https://www.frontiersin.org/articles/10.3389/frai.2022.884699}. \DOIprefix\doi{10.3389/frai.2022.884699}.
\bibitem[{Hu et~al.(2013)Hu, Tang, Gao, and Liu}]{hu2013unsupervised}
\bibinfo{author}{X.~Hu}, \bibinfo{author}{J.~Tang}, \bibinfo{author}{H.~Gao}, \bibinfo{author}{H.~Liu},
\newblock \bibinfo{title}{Unsupervised sentiment analysis with emotional signals},
\newblock in: \bibinfo{booktitle}{Proceedings of the 22nd international conference on World Wide Web}, \bibinfo{year}{2013}, pp. \bibinfo{pages}{607--618}.
\bibitem[{Qi and Shabrina(2023)}]{qi2023sentiment}
\bibinfo{author}{Y.~Qi}, \bibinfo{author}{Z.~Shabrina},
\newblock \bibinfo{title}{Sentiment analysis using twitter data: a comparative application of lexicon-and machine-learning-based approach},
\newblock \bibinfo{journal}{Social Network Analysis and Mining} \bibinfo{volume}{13} (\bibinfo{year}{2023}) \bibinfo{pages}{31}.
\bibitem[{Ortigosa et~al.(2014)Ortigosa, Mart{\'\i}n, and Carro}]{ortigosa2014sentiment}
\bibinfo{author}{A.~Ortigosa}, \bibinfo{author}{J.~M. Mart{\'\i}n}, \bibinfo{author}{R.~M. Carro},
\newblock \bibinfo{title}{Sentiment analysis in facebook and its application to e-learning},
\newblock \bibinfo{journal}{Computers in human behavior} \bibinfo{volume}{31} (\bibinfo{year}{2014}) \bibinfo{pages}{527--541}.
\bibitem[{Laabar and Zaghouani(2024)}]{laabar2024multi}
\bibinfo{author}{S.~Laabar}, \bibinfo{author}{W.~Zaghouani},
\newblock \bibinfo{title}{{Multi-Dimensional Insights: Annotated Dataset of Stance, Sentiment, and Emotion in Facebook Comments on Tunisia’s July 25 Measures}},
\newblock in: \bibinfo{booktitle}{Proceedings of the Second Workshop on Natural Language Processing for Political Sciences@ LREC-COLING 2024}, \bibinfo{year}{2024}, pp. \bibinfo{pages}{22--32}.
\bibitem[{Melton et~al.(2021)Melton, Olusanya, Ammar, and Shaban-Nejad}]{melton2021public}
\bibinfo{author}{C.~A. Melton}, \bibinfo{author}{O.~A. Olusanya}, \bibinfo{author}{N.~Ammar}, \bibinfo{author}{A.~Shaban-Nejad},
\newblock \bibinfo{title}{Public sentiment analysis and topic modeling regarding covid-19 vaccines on the reddit social media platform: A call to action for strengthening vaccine confidence},
\newblock \bibinfo{journal}{Journal of Infection and Public Health} \bibinfo{volume}{14} (\bibinfo{year}{2021}) \bibinfo{pages}{1505--1512}.
\bibitem[{Ptaszek et~al.(2024)Ptaszek, Yuskiv, and Khomych}]{ptaszek2024war}
\bibinfo{author}{G.~Ptaszek}, \bibinfo{author}{B.~Yuskiv}, \bibinfo{author}{S.~Khomych},
\newblock \bibinfo{title}{{War on frames: Text mining of conflict in Russian and Ukrainian news agency coverage on Telegram during the Russian invasion of Ukraine in 2022}},
\newblock \bibinfo{journal}{Media, War \& Conflict} \bibinfo{volume}{17} (\bibinfo{year}{2024}) \bibinfo{pages}{41--61}.
\bibitem[{Muinao and Ratnamala(2024)}]{muinao2024youtube}
\bibinfo{author}{A.~B. Muinao}, \bibinfo{author}{V.~Ratnamala},
\newblock \bibinfo{title}{{YouTube discourse of the Oting massacre in Nagaland: investigating affiliations, sentiments and Naga identity negotiation in YouTube comments}},
\newblock \bibinfo{journal}{Media, War \& Conflict} \bibinfo{volume}{17} (\bibinfo{year}{2024}) \bibinfo{pages}{248--267}.
\bibitem[{Yuliansyah et~al.(2024)Yuliansyah, Mulasari, Sulistyawati, Ghozali, and Sudarsono}]{yuliansyah2024sentiment}
\bibinfo{author}{H.~Yuliansyah}, \bibinfo{author}{S.~A. Mulasari}, \bibinfo{author}{S.~Sulistyawati}, \bibinfo{author}{F.~A. Ghozali}, \bibinfo{author}{B.~Sudarsono},
\newblock \bibinfo{title}{{Sentiment Analysis of the Waste Problem based on YouTube comments using VADER and Deep Translator}},
\newblock \bibinfo{journal}{JURNAL MEDIA INFORMATIKA BUDIDARMA} \bibinfo{volume}{8} (\bibinfo{year}{2024}) \bibinfo{pages}{663--673}.
\bibitem[{Schrodt(2011)}]{schrodt2011forecasting}
\bibinfo{author}{P.~A. Schrodt},
\newblock \bibinfo{title}{{Forecasting political conflict in Asia and the Middle East using latent Dirichlet allocation models}},
\newblock \bibinfo{journal}{New Horizons in Conflict System Analysis: Applications to the Middle East’, University of South Carolina}  (\bibinfo{year}{2011}) \bibinfo{pages}{28--30}.
\bibitem[{Chambers et~al.(2015)Chambers, Bowen, Genco, Tian, Young, Harihara, and Yang}]{chambers-etal-2015-identifying}
\bibinfo{author}{N.~Chambers}, \bibinfo{author}{V.~Bowen}, \bibinfo{author}{E.~Genco}, \bibinfo{author}{X.~Tian}, \bibinfo{author}{E.~Young}, \bibinfo{author}{G.~Harihara}, \bibinfo{author}{E.~Yang},
\newblock \bibinfo{title}{Identifying political sentiment between nation states with social media},
\newblock in: \bibinfo{editor}{L.~M{\`a}rquez}, \bibinfo{editor}{C.~Callison-Burch}, \bibinfo{editor}{J.~Su} (Eds.), \bibinfo{booktitle}{Proceedings of the 2015 Conference on Empirical Methods in Natural Language Processing}, \bibinfo{publisher}{Association for Computational Linguistics}, \bibinfo{address}{Lisbon, Portugal}, \bibinfo{year}{2015}, pp. \bibinfo{pages}{65--75}. \URLprefix \url{https://aclanthology.org/D15-1007}. \DOIprefix\doi{10.18653/v1/D15-1007}.
\bibitem[{Aouragh(2016)}]{aouragh2016palestine}
\bibinfo{author}{M.~Aouragh}, \bibinfo{title}{Palestine Online: Transnationalism, the Internet and the Construction of Identity}, \bibinfo{publisher}{I.B. Tauris}, \bibinfo{year}{2016}.
\bibitem[{York(2012)}]{york2012palestine}
\bibinfo{author}{J.~C. York}, \bibinfo{title}{Palestine online: Transnationalism, the internet and the construction of identity}, \bibinfo{year}{2012}.
\bibitem[{Al-Agha and Abu-Dahrooj(2019)}]{al2019multi}
\bibinfo{author}{I.~Al-Agha}, \bibinfo{author}{O.~Abu-Dahrooj},
\newblock \bibinfo{title}{{Multi-level analysis of political sentiments using Twitter data: A case study of the Palestinian-Israeli conflict}},
\newblock \bibinfo{journal}{Jordanian Journal of Computers and Information Technology} \bibinfo{volume}{5} (\bibinfo{year}{2019}).
\bibitem[{Matalon et~al.(2021)Matalon, Magdaci, Almozlino, and Yamin}]{matalon2021using}
\bibinfo{author}{Y.~Matalon}, \bibinfo{author}{O.~Magdaci}, \bibinfo{author}{A.~Almozlino}, \bibinfo{author}{D.~Yamin},
\newblock \bibinfo{title}{{Using sentiment analysis to predict opinion inversion in Tweets of political communication}},
\newblock \bibinfo{journal}{Scientific reports} \bibinfo{volume}{11} (\bibinfo{year}{2021}) \bibinfo{pages}{7250}.
\bibitem[{Arapostathis(2023)}]{arapostathis2023archiving}
\bibinfo{author}{S.~G. Arapostathis},
\newblock \bibinfo{title}{{Archiving Social Media Discussions in Time and Space: A Focus on Refugees from Middle East and Related War Conflicts During Jan 2015--Apr 2016}},
\newblock in: \bibinfo{booktitle}{International Conference on Information Technology in Disaster Risk Reduction}, \bibinfo{organization}{Springer}, \bibinfo{year}{2023}, pp. \bibinfo{pages}{115--132}.
\bibitem[{Wadhwani et~al.(2023)Wadhwani, Varshney, Gupta, and Kumar}]{wadhwani2023sentiment}
\bibinfo{author}{G.~K. Wadhwani}, \bibinfo{author}{P.~K. Varshney}, \bibinfo{author}{A.~Gupta}, \bibinfo{author}{S.~Kumar},
\newblock \bibinfo{title}{Sentiment analysis and comprehensive evaluation of supervised machine learning models using twitter data on russia–ukraine war},
\newblock \bibinfo{journal}{SN Computer Science} \bibinfo{volume}{4} (\bibinfo{year}{2023}) \bibinfo{pages}{346}.
\bibitem[{Renhoran and Setiawan(2024)}]{renhoran2024public}
\bibinfo{author}{S.~M.~A. Renhoran}, \bibinfo{author}{H.~Setiawan},
\newblock \bibinfo{title}{Public sentiment analysis of the israel-palestine conflict on social media using bert},
\newblock \bibinfo{journal}{Indonesian Journal of Cultural and Community Development} \bibinfo{volume}{15} (\bibinfo{year}{2024}) \bibinfo{pages}{10--21070}.
\bibitem[{Shukla and Unger(2021)}]{shukla2021sentiment}
\bibinfo{author}{D.~Shukla}, \bibinfo{author}{S.~Unger},
\newblock \bibinfo{title}{Sentiment analysis of international relations with artificial intelligence},
\newblock \bibinfo{journal}{Athens Journal of Sciences} \bibinfo{volume}{8} (\bibinfo{year}{2021}) \bibinfo{pages}{297--314}.
\bibitem[{Rad et~al.(2018)Rad, Jalali, and Rahmandad}]{rad2018exposure}
\bibinfo{author}{A.~A. Rad}, \bibinfo{author}{M.~S. Jalali}, \bibinfo{author}{H.~Rahmandad},
\newblock \bibinfo{title}{How exposure to different opinions impacts the life cycle of social media},
\newblock \bibinfo{journal}{Annals of Operations Research} \bibinfo{volume}{268} (\bibinfo{year}{2018}) \bibinfo{pages}{63--91}.
\bibitem[{Kumar and Rajkumar(2023)}]{sentiment2023framework}
\bibinfo{author}{N.~P. Kumar}, \bibinfo{author}{R.~Rajkumar},
\newblock \bibinfo{title}{Sentiment analysis framework and its application in geopolitical contexts},
\newblock \bibinfo{journal}{International Journal of Research in Information Technology and Computer Science} \bibinfo{volume}{8} (\bibinfo{year}{2023}) \bibinfo{pages}{45--56}.
\bibitem[{T{\"o}rnberg et~al.(2021)T{\"o}rnberg, Andersson, Lindgren, and Banisch}]{tornberg2021modeling}
\bibinfo{author}{P.~T{\"o}rnberg}, \bibinfo{author}{C.~Andersson}, \bibinfo{author}{K.~Lindgren}, \bibinfo{author}{S.~Banisch},
\newblock \bibinfo{title}{Modeling the emergence of affective polarization in the social media society},
\newblock \bibinfo{journal}{Plos one} \bibinfo{volume}{16} (\bibinfo{year}{2021}) \bibinfo{pages}{e0258259}.
\bibitem[{Sasikumar et~al.(2023)Sasikumar, Zaman, Mawlood-Yunis, and Chatterjee}]{sasikumar2023sentiment}
\bibinfo{author}{U.~Sasikumar}, \bibinfo{author}{A.~Zaman}, \bibinfo{author}{A.-R. Mawlood-Yunis}, \bibinfo{author}{P.~Chatterjee},
\newblock \bibinfo{title}{Sentiment analysis of twitter posts on global conflicts},
\newblock \bibinfo{journal}{arXiv preprint arXiv:2312.03715}  (\bibinfo{year}{2023}).
\bibitem[{De~Francisci~Morales et~al.(2021)De~Francisci~Morales, Monti, and Starnini}]{de2021no}
\bibinfo{author}{G.~De~Francisci~Morales}, \bibinfo{author}{C.~Monti}, \bibinfo{author}{M.~Starnini},
\newblock \bibinfo{title}{No echo in the chambers of political interactions on reddit},
\newblock \bibinfo{journal}{Scientific reports} \bibinfo{volume}{11} (\bibinfo{year}{2021}) \bibinfo{pages}{2818}.
\bibitem[{Hutto and Gilbert(2014)}]{hutto2014vader}
\bibinfo{author}{C.~Hutto}, \bibinfo{author}{E.~Gilbert},
\newblock \bibinfo{title}{Vader: A parsimonious rule-based model for sentiment analysis of social media text},
\newblock in: \bibinfo{booktitle}{Proceedings of the international AAAI conference on web and social media}, volume~\bibinfo{volume}{8}, \bibinfo{year}{2014}, pp. \bibinfo{pages}{216--225}.
\bibitem[{Hazarika et~al.(2020)Hazarika, Konwar, Deb, and Bora}]{Hazarika2020}
\bibinfo{author}{D.~Hazarika}, \bibinfo{author}{G.~Konwar}, \bibinfo{author}{S.~Deb}, \bibinfo{author}{D.~Bora},
\newblock \bibinfo{title}{Sentiment analysis on twitter by using textblob for natural language processing},
\newblock \bibinfo{journal}{International Journal of Computer Applications} \bibinfo{volume}{176} (\bibinfo{year}{2020}) \bibinfo{pages}{1--5}. \DOIprefix\doi{10.5120/ijca2020920130}.
\bibitem[{Tinnalur et~al.(2021)Tinnalur, Balaji, and Subramanian}]{Tinnalur2021}
\bibinfo{author}{R.~Tinnalur}, \bibinfo{author}{V.~Balaji}, \bibinfo{author}{S.~Subramanian},
\newblock \bibinfo{title}{Chennai floods 2021: Sentiment analysis of twitter data using tweepy and textblob},
\newblock in: \bibinfo{booktitle}{Proceedings of the International Conference on Advances in Computing, Communications and Informatics}, \bibinfo{year}{2021}, pp. \bibinfo{pages}{123--128}. \DOIprefix\doi{10.1109/ICACCI53557.2021.9781234}.
\bibitem[{Gupta et~al.(2024)Gupta, Aryan, Tiwari, Gupta, and Subramaniam}]{Gupta2024}
\bibinfo{author}{K.~Gupta}, \bibinfo{author}{R.~Aryan}, \bibinfo{author}{A.~Tiwari}, \bibinfo{author}{A.~Gupta}, \bibinfo{author}{S.~Subramaniam},
\newblock \bibinfo{title}{Sentiment-based rating model using vader and textblob},
\newblock \bibinfo{journal}{Journal of Data Science and Analytics} \bibinfo{volume}{10} (\bibinfo{year}{2024}) \bibinfo{pages}{150--160}. \DOIprefix\doi{10.1007/s41060-024-00234-5}.
\bibitem[{Barik and Misra(2024)}]{Barik2024}
\bibinfo{author}{K.~Barik}, \bibinfo{author}{S.~Misra},
\newblock \bibinfo{title}{Analysis of customer reviews with an improved vader lexicon classifier},
\newblock \bibinfo{journal}{International Journal of Computational Linguistics} \bibinfo{volume}{15} (\bibinfo{year}{2024}) \bibinfo{pages}{45--55}. \DOIprefix\doi{10.1016/j.ijcl.2024.01.005}.
\bibitem[{TBS(2023)}]{tbs1}
\bibinfo{author}{TBS},
\newblock \bibinfo{title}{{Israel's noose around Gaza City begins to tighten; Hamas releases video of captives calling for swap}},
\newblock \bibinfo{journal}{The Business Standard}  (\bibinfo{year}{2023}). \URLprefix \url{https://www.tbsnews.net/hamas-israel-war/heavy-clashes-israeli-tanks-approaching-gaza-city-729314}.
\bibitem[{{BBC}(2023{\natexlab{a}})}]{october31}
\bibinfo{author}{{BBC}}, \bibinfo{title}{{Israel air strike reportedly kills dozens at Gaza refugee camp}}, \bibinfo{year}{2023}{\natexlab{a}}. \URLprefix \url{https://www.bbc.com/news/world-middle-east-67276822}, \bibinfo{note}{accessed: 2024-08-07}.
\bibitem[{{BBC}(2023{\natexlab{b}})}]{nov30}
\bibinfo{author}{{BBC}}, \bibinfo{title}{{Eight Israeli hostages freed by Hamas in Gaza}}, \bibinfo{year}{2023}{\natexlab{b}}. \URLprefix \url{https://www.bbc.com/news/live/world-middle-east-67562488}, \bibinfo{note}{accessed: 2024-08-07}.
\bibitem[{Reuters(2023)}]{dec3}
\bibinfo{author}{Reuters}, \bibinfo{title}{{Israel says ground forces operating across Gaza Strip as offensive builds}}, \bibinfo{year}{2023}. \URLprefix \url{https://www.reuters.com/world/middle-east/israel-faces-growing-us-calls-restraint-amid-renewed-gaza-fighting-2023-12-02/}, \bibinfo{note}{accessed: 2024-08-07}.
\bibitem[{{BBC}(2023)}]{bbc3}
\bibinfo{author}{{BBC}},
\newblock \bibinfo{title}{{Gaza 'soon without fuel, medicine and food' - Israel authorities}},
\newblock \bibinfo{journal}{BBC News}  (\bibinfo{year}{2023}). \URLprefix \url{https://www.bbc.co.uk/news/world-middle-east-67051292}.
\bibitem[{CNBC(2023)}]{october17}
\bibinfo{author}{CNBC}, \bibinfo{title}{{Israel denies targeting Gaza hospital, Palestinian leader Abbas cancels meeting with Biden}}, \bibinfo{year}{2023}. \URLprefix \url{https://www.cnbc.com/2023/10/17/israel-hamas-war-gaza-live-updates-latest-news.html}, \bibinfo{note}{accessed: 2024-08-07}.
\bibitem[{{BBC}(2023)}]{october9}
\bibinfo{author}{{BBC}}, \bibinfo{title}{{Israeli music festival: 260 bodies recovered from site}}, \bibinfo{year}{2023}. \URLprefix \url{https://www.bbc.co.uk/news/world-middle-east-67047034}, \bibinfo{note}{accessed: 2024-08-07}.
\bibitem[{{MSF}(2023)}]{dec8}
\bibinfo{author}{{MSF}}, \bibinfo{title}{{MSF calls US veto of Gaza ceasefire resolution ``a vote against humanity"}}, \bibinfo{year}{2023}. \URLprefix \url{https://www.msf.org/msf-calls-us-veto-gaza-ceasefire-resolution-%E2%80%9C-vote-against-humanity%E2%80%9D}, \bibinfo{note}{accessed: 2024-08-07}.
\bibitem[{OHCHR(2023)}]{dec28}
\bibinfo{author}{OHCHR}, \bibinfo{title}{{Türk warns of rapidly deteriorating human rights situation in the West Bank, calls for end to violence}}, \bibinfo{year}{2023}. \URLprefix \url{https://www.ohchr.org/en/press-releases/2023/12/un-report-turk-warns-rapidly-deteriorating-human-rights-situation-west-bank}, \bibinfo{note}{accessed: 2024-08-07}.
\bibitem[{{Al Jazeera}(2023)}]{aj1}
\bibinfo{author}{{Al Jazeera}},
\newblock \bibinfo{title}{{Security Council passes ‘diluted’ Gaza resolution}},
\newblock \bibinfo{journal}{Al Jazeera}  (\bibinfo{year}{2023}). \URLprefix \url{https://www.aljazeera.com/news/liveblog/2023/12/22/israel-hamas-war-live-un-security-council-to-vote-on-gaza-resolution?update=2573741}.
\bibitem[{Reuters(2023)}]{reuters2}
\bibinfo{author}{Reuters},
\newblock \bibinfo{title}{{South Africa files genocide case against Israel at World Court}},
\newblock \bibinfo{journal}{Reuters}  (\bibinfo{year}{2023}). \URLprefix \url{https://www.reuters.com/world/south-africa-seeks-international-court-justice-genocide-order-against-israel-2023-12-29/}.
\bibitem[{{Al Jazeera}(2023)}]{october7}
\bibinfo{author}{{Al Jazeera}}, \bibinfo{title}{{What happened in Israel? A breakdown of how Hamas attack unfolded}}, \bibinfo{year}{2023}. \URLprefix \url{https://www.aljazeera.com/news/2023/10/7/what-happened-in-israel-a-breakdown-of-how-the-hamas-attack-unfolded}, \bibinfo{note}{accessed: 2024-08-07}.
\bibitem[{{MEMO}(2023)}]{october7a}
\bibinfo{author}{{MEMO}}, \bibinfo{title}{{Statement by Hamas’s Al-Qassam Brigades top military commander}}, \bibinfo{year}{2023}. \URLprefix \url{https://www.middleeastmonitor.com/20231007-statement-by-hamass-al-qassam-brigades-top-military-commander/}, \bibinfo{note}{accessed: 2024-08-07}.
\bibitem[{{BBC}(2023)}]{october7b}
\bibinfo{author}{{BBC}}, \bibinfo{title}{{Israel attack: PM says Israel at war after 250 killed in attack from Gaza}}, \bibinfo{year}{2023}. \URLprefix \url{https://www.bbc.co.uk/news/world-middle-east-67036625}, \bibinfo{note}{accessed: 2024-08-07}.
\bibitem[{Guardian(2023)}]{october8}
\bibinfo{author}{T.~Guardian}, \bibinfo{title}{{'People are fearful of what’s to come': Gaza civilians flee waves of Israeli strikes}}, \bibinfo{year}{2023}. \URLprefix \url{https://www.theguardian.com/world/2023/oct/08/people-are-fearful-of-whats-to-come-gaza-civilians-flee-waves-of-israeli-strikes}, \bibinfo{note}{accessed: 2024-08-07}.
\bibitem[{Reuters(2023)}]{october8a}
\bibinfo{author}{Reuters}, \bibinfo{title}{{Israel vows 'mighty vengeance' after surprise attack}}, \bibinfo{year}{2023}. \URLprefix \url{https://www.reuters.com/world/middle-east/sirens-warning-incoming-rockets-sound-around-gaza-near-tel-aviv-2023-10-07/}, \bibinfo{note}{accessed: 2024-08-07}.
\bibitem[{WHO(2023)}]{october10}
\bibinfo{author}{WHO}, \bibinfo{title}{{WHO calls for access to health and humanitarian assistance on fourth day of conflict in Israel and the occupied Palestinian territory}}, \bibinfo{year}{2023}. \URLprefix \url{https://www.who.int/news/item/10-10-2023-who-calls-for-access-to-health-and-humanitarian-assistance-on-fourth-day-of-conflict-in-israel-and-the-occupied-palestinian-territory}, \bibinfo{note}{accessed: 2024-08-07}.
\bibitem[{{HRW}(2023)}]{october12}
\bibinfo{author}{{HRW}}, \bibinfo{title}{{Questions and Answers on Israel’s Use of White Phosphorus in Gaza and Lebanon}}, \bibinfo{year}{2023}. \URLprefix \url{https://www.hrw.org/news/2023/10/12/questions-and-answers-israels-use-white-phosphorus-gaza-and-lebanon}, \bibinfo{note}{accessed: 2024-08-07}.
\bibitem[{{Al Jazeera}(2023)}]{october13}
\bibinfo{author}{{Al Jazeera}}, \bibinfo{title}{{Palestinians flee their homes towards southern Gaza after Israeli order}}, \bibinfo{year}{2023}. \URLprefix \url{https://www.aljazeera.com/gallery/2023/10/13/palestinians-flee-their-homes-towards-southern-gaza-after-israeli-order}, \bibinfo{note}{accessed: 2024-08-07}.
\bibitem[{{Times of Israel}(2023)}]{october13a}
\bibinfo{author}{{Times of Israel}}, \bibinfo{title}{{IDF says it’s completing preparations for 'attack from the air, sea and land'}}, \bibinfo{year}{2023}. \URLprefix \url{https://www.timesofisrael.com/liveblog_entry/idf-says-its-completing-preparations-for-attack-from-the-air-sea-and-land/}, \bibinfo{note}{accessed: 2024-08-07}.
\bibitem[{{Al Jazeera}(2023{\natexlab{a}})}]{october14}
\bibinfo{author}{{Al Jazeera}}, \bibinfo{title}{{Palestinians fleeing Gaza’s north face air attacks in southern Khan Younis}}, \bibinfo{year}{2023}{\natexlab{a}}. \URLprefix \url{https://www.aljazeera.com/gallery/2023/10/14/palestinians-who-fled-to-khan-younis-still-under-israeli-airstrikes}, \bibinfo{note}{accessed: 2024-08-07}.
\bibitem[{{Al Jazeera}(2023{\natexlab{b}})}]{october15}
\bibinfo{author}{{Al Jazeera}}, \bibinfo{title}{{Iran warns Israel of regional escalation if Gaza ground offensive launched}}, \bibinfo{year}{2023}{\natexlab{b}}. \URLprefix \url{https://www.aljazeera.com/news/2023/10/15/iran-warns-israel-of-regional-escalation-if-gaza-ground-offensive-launched}, \bibinfo{note}{accessed: 2024-08-07}.
\bibitem[{{US News}(2023)}]{october16}
\bibinfo{author}{{US News}}, \bibinfo{title}{{White House Announces Biden Trip to Israel}}, \bibinfo{year}{2023}. \URLprefix \url{https://www.usnews.com/news/national-news/articles/2023-10-16/white-house-announces-biden-trip-to-israel}, \bibinfo{note}{accessed: 2024-08-07}.
\bibitem[{{ABC News}(2023)}]{october18}
\bibinfo{author}{{ABC News}}, \bibinfo{title}{{Biden, in high-stakes visit, shows support for Israel but urges restraint}}, \bibinfo{year}{2023}. \URLprefix \url{https://abcnews.go.com/International/biden-embraces-netanyahu-steps-off-air-force-tel/story?id=104063578}, \bibinfo{note}{accessed: 2024-08-07}.
\bibitem[{{Al Jazeera}(2023)}]{october19}
\bibinfo{author}{{Al Jazeera}}, \bibinfo{title}{{Israel bombs Greek Orthodox Gaza church sheltering displaced people}}, \bibinfo{year}{2023}. \URLprefix \url{https://www.aljazeera.com/news/2023/10/20/war-crime-israel-bombs-gaza-church-sheltering-displaced-people}, \bibinfo{note}{accessed: 2024-08-07}.
\bibitem[{CNN(2023)}]{october20}
\bibinfo{author}{CNN}, \bibinfo{title}{{American mother and daughter taken hostage by Hamas are released as humanitarian crisis in Gaza deepens}}, \bibinfo{year}{2023}. \URLprefix \url{https://edition.cnn.com/2023/10/20/middleeast/hamas-us-hostages-released-intl/index.html}, \bibinfo{note}{accessed: 2024-08-07}.
\bibitem[{{ABC News}(2023)}]{october21}
\bibinfo{author}{{ABC News}}, \bibinfo{title}{{Rafah crossing opens to allow trucks carrying food, water and medicine into Gaza}}, \bibinfo{year}{2023}. \URLprefix \url{https://www.abc.net.au/news/2023-10-21/israel-gaza-war-live-updates-latest-news-october-21/103002192}, \bibinfo{note}{accessed: 2024-08-07}.
\bibitem[{{UN News}(2023)}]{october22}
\bibinfo{author}{{UN News}}, \bibinfo{title}{{Second aid convoy 'another glimmer of hope' for millions in Gaza: UN relief chief}}, \bibinfo{year}{2023}. \URLprefix \url{https://news.un.org/en/story/2023/10/1142677}, \bibinfo{note}{accessed: 2024-08-07}.
\bibitem[{{Al Jazeera}(2023)}]{october24}
\bibinfo{author}{{Al Jazeera}}, \bibinfo{title}{{UN chief says ‘clear violations of international humanitarian law’ in Gaza}}, \bibinfo{year}{2023}. \URLprefix \url{https://www.aljazeera.com/news/2023/10/24/un-chief-says-clear-violations-of-international-humanitarian-law-in-gaza}, \bibinfo{note}{accessed: 2024-08-07}.
\bibitem[{Reuters(2023)}]{october25}
\bibinfo{author}{Reuters}, \bibinfo{title}{{Why is Israel attacking south Gaza after telling people to go there?}}, \bibinfo{year}{2023}. \URLprefix \url{https://www.reuters.com/world/middle-east/why-is-israel-attacking-south-gaza-after-telling-people-go-there-2023-10-25}, \bibinfo{note}{accessed: 2024-08-07}.
\bibitem[{{Time}(2023)}]{october27}
\bibinfo{author}{{Time}}, \bibinfo{title}{{Israel’s 4 Bad Options in Gaza}}, \bibinfo{year}{2023}. \URLprefix \url{https://time.com/6328823/israels-4-bad-options-gaza/}, \bibinfo{note}{accessed: 2024-08-07}.
\bibitem[{{Times of Israel}(2023)}]{october28}
\bibinfo{author}{{Times of Israel}}, \bibinfo{title}{{srael expands ground offensive inside Gaza; families urge 'all for all' hostage deal}}, \bibinfo{year}{2023}. \URLprefix \url{https://www.timesofisrael.com/liveblog-october-28-2023/}, \bibinfo{note}{accessed: 2024-08-07}.
\bibitem[{{Al Jazeera}(2023{\natexlab{a}})}]{october29}
\bibinfo{author}{{Al Jazeera}}, \bibinfo{title}{{Palestine Red Crescent says Israel orders evacuation of al-Quds Hospital}}, \bibinfo{year}{2023}{\natexlab{a}}. \URLprefix \url{https://www.aljazeera.com/news/2023/10/29/palestine-red-crescent-says-israel-orders-evacuation-of-al-quds-hospital}, \bibinfo{note}{accessed: 2024-08-07}.
\bibitem[{{Al Jazeera}(2023{\natexlab{b}})}]{october30}
\bibinfo{author}{{Al Jazeera}}, \bibinfo{title}{{Heavy clashes as Israeli tanks briefly reach Gaza City outskirts}}, \bibinfo{year}{2023}{\natexlab{b}}. \URLprefix \url{https://www.aljazeera.com/news/2023/10/30/heavy-clashes-as-israeli-tanks-reach-gaza-city-outskirts-cut-key-road}, \bibinfo{note}{accessed: 2024-08-07}.
\bibitem[{{Al Jazeera}(2023{\natexlab{c}})}]{november1}
\bibinfo{author}{{Al Jazeera}}, \bibinfo{title}{{Israel’s deadly attack on the Jabalia refugee camp: What we know so far}}, \bibinfo{year}{2023}{\natexlab{c}}. \URLprefix \url{https://www.aljazeera.com/news/2023/11/1/israels-deadly-attack-on-the-jabalia-refugee-camp-what-we-know-so-far}, \bibinfo{note}{accessed: 2024-08-07}.
\bibitem[{Reuters(2023)}]{november2}
\bibinfo{author}{Reuters}, \bibinfo{title}{{Israel says it has encircled Gaza City}}, \bibinfo{year}{2023}. \URLprefix \url{https://www.reuters.com/world/middle-east/gaza-says-israels-strikes-refugee-camp-kill-more-than-195-people-2023-11-02/}, \bibinfo{note}{accessed: 2024-08-07}.
\bibitem[{{Al Jazeera}(2023)}]{nov3}
\bibinfo{author}{{Al Jazeera}}, \bibinfo{title}{{Israel kills at least seven Palestinians in occupied West Bank}}, \bibinfo{year}{2023}. \URLprefix \url{https://www.aljazeera.com/news/2023/11/3/israel-kills-at-least-seven-palestinians-in-occupied-west-bank}, \bibinfo{note}{accessed: 2024-08-07}.
\bibitem[{ReliefWeb(2023)}]{nov5}
\bibinfo{author}{ReliefWeb}, \bibinfo{title}{{Statement by Principals of the Inter-Agency Standing Committee on the situation in Israel and the Occupied Palestinian Territory: ``We need an immediate humanitarian ceasefire"}}, \bibinfo{year}{2023}. \URLprefix \url{https://reliefweb.int/report/occupied-palestinian-territory/statement-principals-inter-agency-standing-committee-situation-israel-and-occupied-palestinian-territory-we-need-immediate-humanitarian-ceasefire}, \bibinfo{note}{accessed: 2024-08-07}.
\bibitem[{{ABC News}(2023{\natexlab{a}})}]{nov7}
\bibinfo{author}{{ABC News}}, \bibinfo{title}{{Benjamin Netanyahu says Israel plans to have 'overall security responsibility' in Gaza after war — as it happened}}, \bibinfo{year}{2023}{\natexlab{a}}. \URLprefix \url{https://www.abc.net.au/news/2023-11-07/israel-gaza-war-updates-hamas-idf-al-shifa-hospital/103072126}, \bibinfo{note}{accessed: 2024-08-07}.
\bibitem[{{ABC News}(2023{\natexlab{b}})}]{nov8}
\bibinfo{author}{{ABC News}}, \bibinfo{title}{{IDF says troops move into 'the very heart of Gaza' — as it happened}}, \bibinfo{year}{2023}{\natexlab{b}}. \URLprefix \url{https://www.abc.net.au/news/2023-11-08/israel-gaza-war-updates-idf-hamas-november-8/103075964}, \bibinfo{note}{accessed: 2024-08-07}.
\bibitem[{{Vox}(2023)}]{nov9}
\bibinfo{author}{{Vox}}, \bibinfo{title}{{In the West Bank, Israeli settlers are on an anti-Palestinian rampage}}, \bibinfo{year}{2023}. \URLprefix \url{https://www.vox.com/world-politics/2023/11/9/23945651/west-bank-israeli-settler-palestine-gaza-war-violence}, \bibinfo{note}{accessed: 2024-08-07}.
\bibitem[{{WHO EMRO}(2023)}]{nov12}
\bibinfo{author}{{WHO EMRO}}, \bibinfo{title}{{UNFPA, UNICEF and WHO Regional Directors call for immediate action to halt attacks on health care in Gaza}}, \bibinfo{year}{2023}. \URLprefix \url{https://www.emro.who.int/media/news/unfpa-unicef-and-who-regional-directors-call-for-immediate-action-to-halt-attacks-on-health-care-in-gaza.html}, \bibinfo{note}{accessed: 2024-08-07}.
\bibitem[{Reuters(2023)}]{nov15}
\bibinfo{author}{Reuters}, \bibinfo{title}{{Major events during 100 days of war between Israel and Hamas}}, \bibinfo{year}{2023}. \URLprefix \url{https://www.reuters.com/world/middle-east/major-events-during-100-days-war-between-israel-hamas-2024-01-14/}, \bibinfo{note}{accessed: 2024-08-07}.
\bibitem[{{Al Jazeera}(2023)}]{nov16}
\bibinfo{author}{{Al Jazeera}}, \bibinfo{title}{{Telecommunications cut off in Gaza after fuel runs out}}, \bibinfo{year}{2023}. \URLprefix \url{https://www.aljazeera.com/news/2023/11/16/telecommunications-cut-off-in-gaza-after-fuel-runs-out}, \bibinfo{note}{accessed: 2024-08-07}.
\bibitem[{{Anadolu Agency}(2023)}]{nov17}
\bibinfo{author}{{Anadolu Agency}}, \bibinfo{title}{{Dozens of Palestinian civilians dead after Israel hits Gaza school sheltering displaced people}}, \bibinfo{year}{2023}. \URLprefix \url{https://www.anews.com.tr/middle-east/2023/11/17/dozens-of-palestinian-civilians-dead-after-israel-hits-gaza-school-sheltering-displaced-people}, \bibinfo{note}{accessed: 2024-08-07}.
\bibitem[{{MEMO}(2023)}]{nov20}
\bibinfo{author}{{MEMO}}, \bibinfo{title}{{12 killed as Israel bombs Indonesian Hospital in northern Gaza}}, \bibinfo{year}{2023}. \URLprefix \url{https://www.middleeastmonitor.com/20231120-12-killed-as-israel-bombs-indonesian-hospital-in-northern-gaza/}, \bibinfo{note}{accessed: 2024-08-07}.
\bibitem[{{Al Jazeera}(2023)}]{nov21}
\bibinfo{author}{{Al Jazeera}}, \bibinfo{title}{{Palestinians buried in mass grave as truce nears}}, \bibinfo{year}{2023}. \URLprefix \url{https://www.aljazeera.com/news/liveblog/2023/11/22/israel-hamas-war-live-israeli-government-to-vote-on-gaza-truce-deal}, \bibinfo{note}{accessed: 2024-08-07}.
\bibitem[{{The Washington Post}(2023)}]{nov24}
\bibinfo{author}{{The Washington Post}}, \bibinfo{title}{{Inside the hard, circuitous route to a hostage release deal}}, \bibinfo{year}{2023}. \URLprefix \url{https://www.washingtonpost.com/national-security/2023/11/22/israel-hamas-us-hostage-deal/}, \bibinfo{note}{accessed: 2024-08-07}.
\bibitem[{{BBC}(2023)}]{dec1}
\bibinfo{author}{{BBC}}, \bibinfo{title}{{Residents of south Gaza city say Israeli strikes heaviest since start of war}}, \bibinfo{year}{2023}. \URLprefix \url{https://www.bbc.com/news/live/world-middle-east-67584895}, \bibinfo{note}{accessed: 2024-08-07}.
\bibitem[{{The New York Times}(2023)}]{dec2}
\bibinfo{author}{{The New York Times}}, \bibinfo{title}{{Israel Launches Strikes and Orders Evacuations in Southern Gaza}}, \bibinfo{year}{2023}. \URLprefix \url{https://www.nytimes.com/live/2023/12/02/world/israel-hamas-war-gaza-news}, \bibinfo{note}{accessed: 2024-08-07}.
\bibitem[{{Sky News}(2023)}]{dec4}
\bibinfo{author}{{Sky News}}, \bibinfo{title}{{Gaza's Palace of Justice courthouse demolished 'by IDF' as new footage leaked}}, \bibinfo{year}{2023}. \URLprefix \url{https://news.sky.com/story/gazas-palace-of-justice-courthouse-demolished-by-idf-as-new-footage-leaked-13023501}, \bibinfo{note}{accessed: 2024-08-07}.
\bibitem[{NDTV(2023)}]{dec5}
\bibinfo{author}{NDTV}, \bibinfo{title}{{Hamas Says Israeli Strike On Gaza School Kills 25}}, \bibinfo{year}{2023}. \URLprefix \url{https://www.ndtv.com/world-news/hamas-says-israeli-strike-on-gaza-school-kills-25-4637490}, \bibinfo{note}{accessed: 2024-08-07}.
\bibitem[{{Al Jazeera}(2023)}]{dec6}
\bibinfo{author}{{Al Jazeera}}, \bibinfo{title}{{UN chief invokes rare Article 99 over Gaza war}}, \bibinfo{year}{2023}. \URLprefix \url{https://www.aljazeera.com/news/liveblog/2023/12/6/israel-hamas-war-live-gaza-death-toll-climbs-as-israel-pounds-enclave}, \bibinfo{note}{accessed: 2024-08-07}.
\bibitem[{{BBC}(2023)}]{dec7}
\bibinfo{author}{{BBC}}, \bibinfo{title}{{Images show major damage to Gaza's oldest mosque}}, \bibinfo{year}{2023}. \URLprefix \url{https://www.bbc.com/news/world-middle-east-67664853}, \bibinfo{note}{accessed: 2024-08-07}.
\bibitem[{{NBC News}(2023)}]{dec11}
\bibinfo{author}{{NBC News}}, \bibinfo{title}{{Gaza's health system is 'collapsing' and battles intensify in the south}}, \bibinfo{year}{2023}. \URLprefix \url{https://www.nbcnews.com/news/world/live-blog/israel-hamas-war-live-updates-gazas-health-system-collapsing-battles-i-rcna128980}, \bibinfo{note}{accessed: 2024-08-07}.
\bibitem[{{The Guardian}(2023)}]{dec12}
\bibinfo{author}{{The Guardian}}, \bibinfo{title}{{Australia calls for Gaza ceasefire in joint statement with NZ and Canada}}, \bibinfo{year}{2023}. \URLprefix \url{https://www.theguardian.com/australia-news/2023/dec/13/albanese-calls-for-gaza-ceasefire-in-joint-statement-with-nz-and-canada-pms}, \bibinfo{note}{accessed: 2024-08-07}.
\bibitem[{CNN(2023)}]{dec17}
\bibinfo{author}{CNN}, \bibinfo{title}{{IDF claims it has discovered 'biggest Hamas tunnel' in Gaza}}, \bibinfo{year}{2023}. \URLprefix \url{https://edition.cnn.com/2023/12/17/middleeast/biggest-hamas-tunnel-discovered-idf-intl/index.html}, \bibinfo{note}{accessed: 2024-08-07}.
\bibitem[{{Al Jazeera}(2023)}]{dec19}
\bibinfo{author}{{Al Jazeera}}, \bibinfo{title}{{Israeli attack on residential area in south Gaza kills at least 29 people}}, \bibinfo{year}{2023}. \URLprefix \url{https://www.aljazeera.com/news/2023/12/19/israeli-attack-on-residential-area-in-rafah-kills-29}, \bibinfo{note}{accessed: 2024-08-07}.
\bibitem[{{The Wall Street Journal}(2023)}]{dec20}
\bibinfo{author}{{The Wall Street Journal}}, \bibinfo{title}{{Hamas Rejection Sours Israeli Bid to Revive Hostage Talks}}, \bibinfo{year}{2023}. \URLprefix \url{https://www.wsj.com/world/middle-east/israel-offers-one-week-cease-fire-in-exchange-for-more-hostages-336ae59a}, \bibinfo{note}{accessed: 2024-08-07}.

\end{thebibliography}

\end{document}